\documentclass[10pt,conference]{IEEEtran}

\usepackage{cite}
\usepackage{balance}
\usepackage{amsmath,amssymb,amsfonts,amsthm}
\theoremstyle{definition}
\newtheorem{definition}{Definition}[section]
\usepackage{graphicx}
\usepackage{textcomp}
\usepackage{xcolor}
\usepackage{booktabs}
\def\BibTeX{{\rm B\kern-.05em{\sc i\kern-.025em b}\kern-.08em
    T\kern-.1667em\lower.7ex\hbox{E}\kern-.125emX}}

\usepackage{subcaption}
\usepackage[vlined,figure,linesnumbered]{algorithm2e}

\usepackage{tikz}
\usepackage{pgfplots}

\definecolor{dkgreen}{rgb}{0,0.5,0}
\definecolor{dkred}{rgb}{0.5,0,0}
\definecolor{dkgray}{rgb}{0.3,0.3,0.3}

\usepackage{multicol}
\usepackage{listings}
\newcommand{\myca}{C5$_{\textbf{i}}$}
\newcommand{\tool}{\texttt{GenTree}}

\begin{document}

\title{{\tool}: Using Decision Trees to Learn Interactions for Configurable Software}

\author{\IEEEauthorblockN{KimHao Nguyen and ThanhVu Nguyen}
\IEEEauthorblockA{
\textit{University of Nebraska-Lincoln, USA}\\
\{kdnguyen,tnguyen\}@cse.unl.edu}
}

\maketitle
\begin{abstract}
  Modern software systems are increasingly designed to be highly configurable, which increases flexibility but can make programs harder to develop, test, and analyze, e.g., how configuration options are set to reach certain locations, what characterizes the configuration space of an interesting or buggy program behavior?
We introduce {\tool}, a new dynamic analysis that automatically learns a program's \emph{interactions}---logical formulae that describe how configuration option settings map to code coverage.
  {\tool} uses an iterative refinement approach that runs the program under a small sample of configurations to obtain coverage data; uses a custom classifying algorithm on these data to build decision trees representing interaction candidates; and then analyzes the trees to generate new configurations to further refine the trees and interactions in the next iteration.
  Our experiments on 17 configurable systems spanning 4 languages show that {\tool} efficiently finds precise interactions using a tiny fraction of the configuration space.

\end{abstract}

\section{Introduction}
Modern software systems are increasingly designed to be configurable.
This has many benefits, but also significantly complicates tasks such as testing, debugging, and analysis due to the number of configurations that can be exponentially large---in the worst case, every combination of option settings can lead to a distinct behavior.
This software \emph{configuration space explosion} presents real challenges to software developers.
It makes testing and debugging more difficult as faults are often visible under only specific combinations of configuration options.
It also causes a challenge to static analyses because configurable systems often have huge configuration spaces and use libraries and native code that are difficult to reason about.

Existing works on highly-configurable systems~\cite{reisner2010using,song:icse12,song:2014aa,nguyen2016igen} showed that we can automatically find \emph{interactions} to concisely describe the configuration space of the system.
These works focus on program coverage (but can be generalized to arbitrary program behaviors) and define an interaction for a location as a logically weakest formula over configuration options such that any configuration satisfying that formula would cover that location.
These works showed that interactions are useful to understand the configurations of the system, e.g., determine what configuration settings cover a given location;  determine what locations a given interaction covers; find important options, and compute a minimal set of configurations to achieve certain coverage; etc.
In the software production line community, feature interactions and presence conditions ({\S}\ref{sec:related}) are similar to interactions and has led to many automated \emph{configuration-aware} testing techniques to debug functional (e.g., bug triggers, memory leaks) and non-functional (e.g., performance anomalies, power consumption) behaviors.
Interactions also help reverse engineering and impact analysis~\cite{She:2011:REF:1985793.1985856,Berger:2010:VMR:1858996.1859010}, and even in the bioinformatics systems for aligning and analyzing DNA sequences~\cite{cashman2018navigating}.

These interaction techniques are promising, but have several limitations.
The symbolic execution work in~\cite{reisner2010using} does not scale to large systems, even when being restricted to configuration options with a small number of values (e.g., boolean); needs user-supplied models (mocks) to represent libraries, frameworks, and native code; and is language-specific (C programs).
iTree~\cite{song:icse12,song:2014aa} uses decision trees to generate configurations to maximize coverage, but achieves very few and imprecise interactions.
Both of these works only focus on interactions that can be represented as purely conjunctive formulae.

The iGen interaction work~\cite{nguyen2016igen} adopts the iterative refinement approach often used to find program preconditions and invariants (e.g.,~\cite{sharma2013data,garg2014ice,garg2016learning,nguyen2017syminfer}).
This approach learns candidate invariants from program execution traces and uses an oracle (e.g., a static checker) to check the candidates.
When the candidate invariants are incorrect, the oracle returns  counterexample traces that the dynamic inference engine can use to infer more accurate invariants.
iGen adapts this iterative algorithm to finding interactions, but avoids static checking, which has limitations similar to symbolic execution as mentioned above.
Instead, iGen modifies certain parts of the candidate interaction to generate new configurations and run them to test the candidate. Configurations that ``break'' the interaction are counterexamples used to improve that interaction in the next iteration.
However, to effectively test interactions and generate counterexample configurations, iGen is restricted to learning interactions under specific forms (purely conjunctive, purely disjunctive, and specific mixtures of the two) and thus cannot capture complex interactions in real-world systems  ({\S}\ref{sec:eval}).

\begin{figure*}[h!]
  \begin{minipage}[b]{0.32\linewidth}
    \centering
    \includegraphics[width=0.9\linewidth]{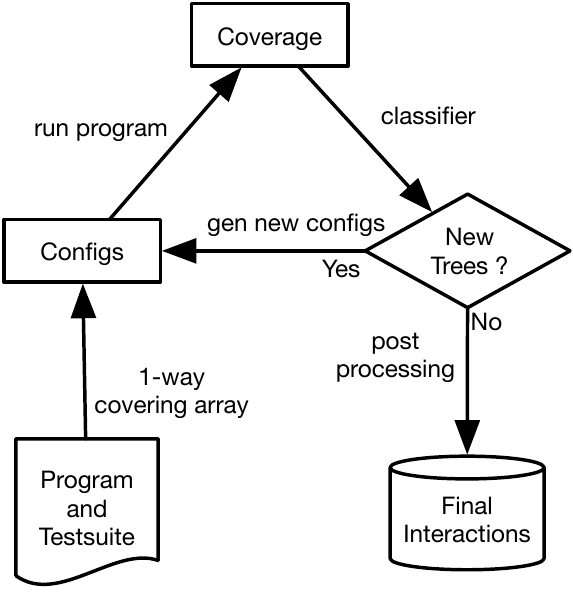}
    \vspace{1em}
    \caption{{\tool} overview}
    \label{fig:overview}
  \end{minipage}
  \hfill
  \begin{minipage}[b]{0.60\linewidth}
    \centering
  \newcommand{\excmt}{\color{dkgreen}}
  \newcommand{\eq}{\equiv}
  \begin{lstlisting}[numbers=none,multicols=2]
//9 configuration options:
//$\excmt s,t,u,v ~(bool)$; $\excmt a,b,c,d,e \in \{0,1,2\}$

printf("$L0$\n"); // $\excmt True$

if ($a \equiv 1 \lor b \equiv 2$) {
  printf("L1\n"); //$\excmt a \equiv 1 \lor b \equiv 2$
}
else if ($c \equiv 0 \wedge d \equiv 1$) {
  //$\excmt a\in\{0,2\} \land b\in\{0,1\} \land c\eq 0 \land d\eq 1 $
  printf("L2\n");
}

if($u \land v$) {
  printf("$L3$\n"); //$\excmt u \land v$
  return;
}
else {
  printf("$L4$\n"); //$\excmt \bar u \lor \bar v$
  if ($s \land e \equiv 2$){
    //$\excmt s \land e\eq 2 \land (\bar u \lor \bar v)$
    printf("$L5$\n");
    return;
  }
}

//$\excmt (\bar s \lor e\in\{0,1\}) \land (\bar u \lor \bar v)$
printf("$L6$\n");

if($e \equiv 2$) {
  //$\excmt \bar s \land e\eq 2 \land (\bar u \lor \bar v)$
  printf("$L7$\n"); 
  if($u \vee v$) {
    //$\excmt \bar s \land e\eq 2 \land ((u \land \bar v) \lor (\bar u \land v))$
    printf("$L8$\n");
  }
}
\end{lstlisting}
\vspace{1em}
\caption{A program having nine locations $L0–L8$ annotated with interactions}
\label{fig:exprog}
\end{minipage}
\end{figure*}

In this paper, we introduce {\tool}, a new dynamic interaction inference technique inspired by the iterative invariant refinement algorithm and iGen.
Figure~\ref{fig:overview} gives an overview of {\tool}.
First, {\tool} creates an initial set of configurations and runs the program to obtain (location) coverage.
Then for each covered location $l$, {\tool} builds a decision tree, which represents a candidate interaction, from the configurations that do and do not cover $l$.

Because {\tool} works with just a sample of all configurations, the decision trees representing candidate interactions may be imprecise.
To refine these trees, {\tool} analyzes them to generate new configurations.
In the next iteration, these configurations may provide the necessary data to invalidate the current trees (i.e., counterexamples) and build more precise trees, which correspond to better interactions.
This process continues until we obtain no new coverage or trees for several consecutive iterations, at which point {\tool} returns the final set of interactions.

The design of {\tool} helps mitigate several limitations of existing works.
By using dynamic analysis, {\tool} is language agnostic and supports complex programs  (e.g., those using third party libraries) that might be difficult for static analyses.
By considering only small configuration samples, {\tool} is efficient and scales well to large programs.
By integrating with iterative refinement, {\tool} generates small sets of useful configurations to gradually improve its results.
By using decision trees, {\tool} supports expressive interactions representing arbitrary boolean formulae and allows for generating effective counterexample configurations.
Finally, by using a classification algorithm customized for interactions, {\tool} can build trees from small data samples to represent accurate interactions.

We evaluated \tool\ on 17 programs in C, Python, Perl, and OCaml having configuration spaces containing $1024$ to $3.5 \times 10^{14}$ configurations.
We found that interaction results from {\tool} are precise, i.e., similar to what {\tool} would produce if it inferred interactions from all possible configurations.
We also found that {\tool} scales well to programs with many options because it only explores a small fraction of the large configuration spaces.
We examined {\tool}'s results and found that they confirmed several observations made by prior work (e.g., conjunctive interactions are common but disjunctive and mixed interactions are still important for coverage; and enabling options, which must be set in a certain way to cover most locations, are common). 
We also observed that complex interactions supported by {\tool} but not from prior works cover a non-trivial number of locations and are critical to understand the program behaviors at these locations.

In summary, this paper makes the following contributions:
(i) we introduce a new iterative refinement algorithm that uses decision trees to represent and refine program interactions;
(ii) we present a decision tree classification algorithm optimized for interaction discovery;
(iii) we implement these ideas in the {\tool} tool and make it freely available;
and (iv) we evaluate {\tool} on programs written in various languages and analyze its results to find interesting configuration properties.
{\tool} and all benchmark data are available at~\cite{toolwebsite}.

\section{Illustration}
\label{sec:overview}

\newcommand{\exloc}{$L8$}

\begin{figure*}[h]
  \centering
  \begin{minipage}{1\linewidth}
    \begin{minipage}{0.65\linewidth}
      \[
        \small
        \begin{array}{c|rrrrrrrrr|l}
          \text{config} & s & t & u & v & a & b & c & d & e& \text{cov } (L)\\
          \midrule
          c_1 & 1 & 1 & 0 & 0 & 0 & 1 & 2 & 1 & 0 & 0,4,6 \\
          c_2 & 0 & 1 & 1 & 0 & 2 & 0 & 0 & 2 & 2 & 0,4,6,7,8 \\
          c_3 & 1 & 0 & 1 & 1 & 1 & 2 & 1 & 0 & 1 & 0,1,3 \\
        \end{array}
      \]
    \end{minipage}
    \hfill
    \begin{minipage}{0.34\linewidth}
      \centering
      \includegraphics[width=0.5\linewidth]{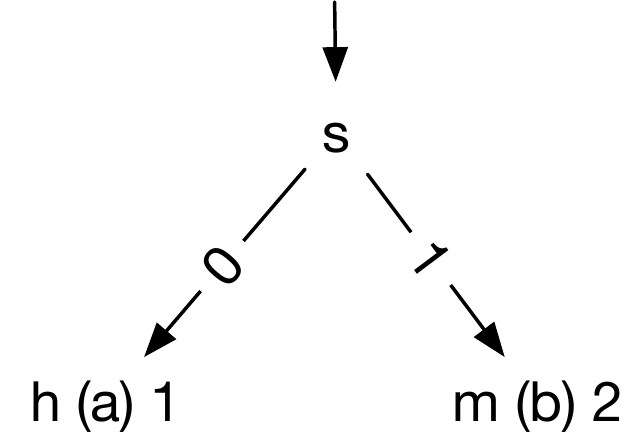}
    \end{minipage}
    \caption{Initial configurations using 1-way covering array and the decision tree for {\exloc} built from these configurations.
      The label at the leaf, e.g., h (a) 1, indicates the classification (hit or miss) of the path from the root to this leaf, the (name) of the path, and the number of configurations used for this classification}
  \label{fig:f1}    
  \end{minipage}
  \begin{minipage}{1\linewidth}
    \begin{minipage}{0.65\linewidth}
      \[
        \small
        \begin{array}{c|rrrrrrrrr|l}
          \text{config} & s & t & u & v & a & b & c & d & e & \text{cov } (L)\\
          \midrule
          \underline{c_4} & 0 & 1 & 1 & 1 & 1 & 1 & 0 & 1 &  0 & 0, 1, 3  \\
          \underline{c_5} & 0 & 0 & 0 & 0 & 0 & 2 & 2 & 0 &  1 & 0, 1, 4, 6  \\
          \underline{c_6} & 0 & 1 & 0 & 0 & 2 & 0 & 1 & 2 &  2 & 0, 4, 6, 7  \\
          c_7 & 0 & 1 & 0 & 1 & 0 & 0 & 1 & 2 &  2 & 0, 4, 6, 7, 8 \\
        \end{array}
      \]
    \end{minipage}
    \hfill
    \begin{minipage}{0.34\linewidth}
      \centering
      \includegraphics[width=0.7\linewidth]{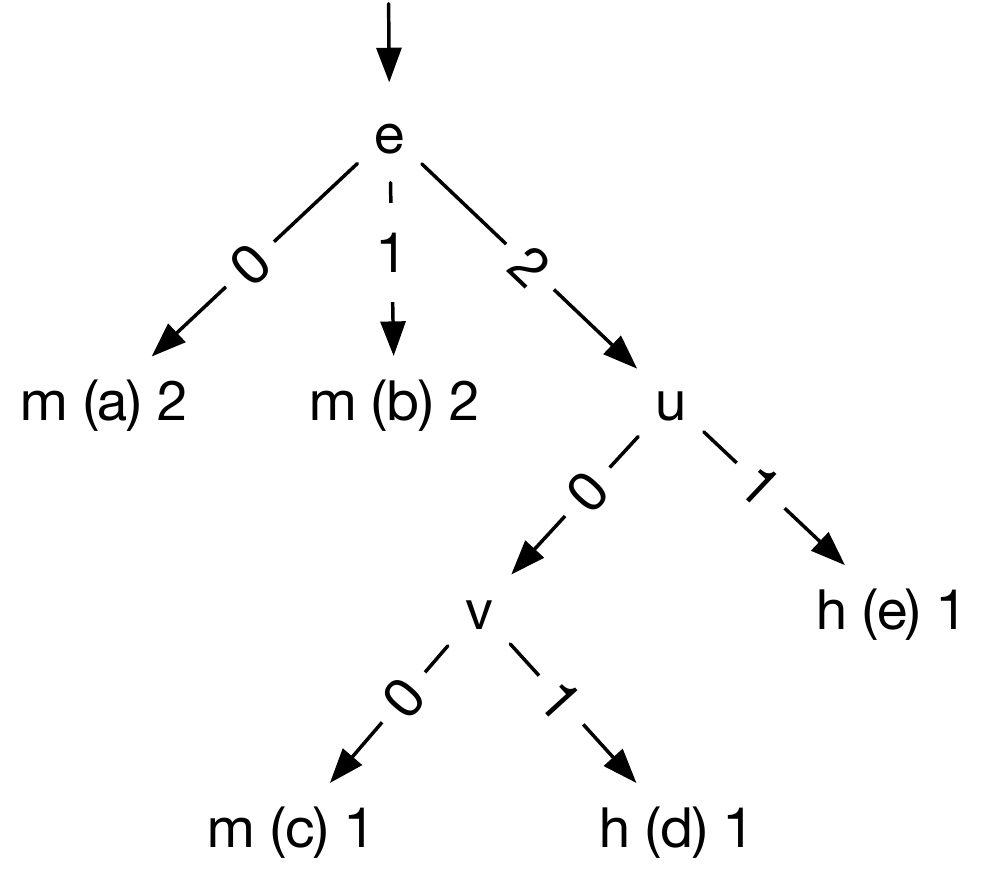}
    \end{minipage}
    \caption{New configurations created from path (a) in the tree in Figure~\ref{fig:f1} (counterexamples are \underline{underlined}) and new decision tree for {\exloc}.}
    \label{fig:f2}    
  \end{minipage}
  \\
  \begin{minipage}{1\linewidth}
    \begin{minipage}{0.65\linewidth}
      \centering
      \[
        \small
        \begin{array}{c|rrrrrrrrr|l|c}
          \text{config} & s & t & u & v & a & b & c & d & e & \text{cov } (L)& \text{path in Fig.~\ref{fig:f2}} \\
          \midrule
          c_{8}   & 0 & 0 & 0 & 0 & 0 & 0 & 2 & 2 & 2 &  0, 4, 6, 7 	& c  \\
          c_{9}   & 1 & 1 & 0 & 0 & 2 & 1 & 1 & 0 & 2 &  0, 4, 5 		& c  \\
          c_{10} & 1 & 0 & 0 & 0 & 1 & 2 & 0 & 1 & 2 &  0,1, 4, 5 	& c  \\
          c_{11} & 0 & 0 & 0 & 1 & 2 & 2 & 1 & 0 & 2 &  0,1, 4, 6, 7, 8       & d  \\
          c_{12} & 0 & 0 & 0 & 1 & 1 & 0 & 2 & 1 & 2 &  0,1, 4, 6, 7, 8       & d  \\
          \underline{c_{13}} & 1 & 1 & 0 & 1 & 0 & 1 & 0 & 2 & 2 &  0, 4, 5& d  \\
          \underline{c_{14}} & 0 & 1 & 1 & 1 & 1 & 2 & 1 & 0 & 2 &  0, 1, 3& e  \\
          \underline{c_{15}} & 1 & 0 & 1 & 0 & 0 & 0 & 2 & 1 & 2 &  0, 4,5& e  \\
          \underline{c_{16}} & 1 & 0 & 1 & 0 & 2 & 1 & 0 & 2 & 2 &  0, 4,5& e  \\
        \end{array}
      \]
    \end{minipage}
    \hfill
    \begin{minipage}{0.34\linewidth}
      \centering
      \includegraphics[width=0.8\linewidth]{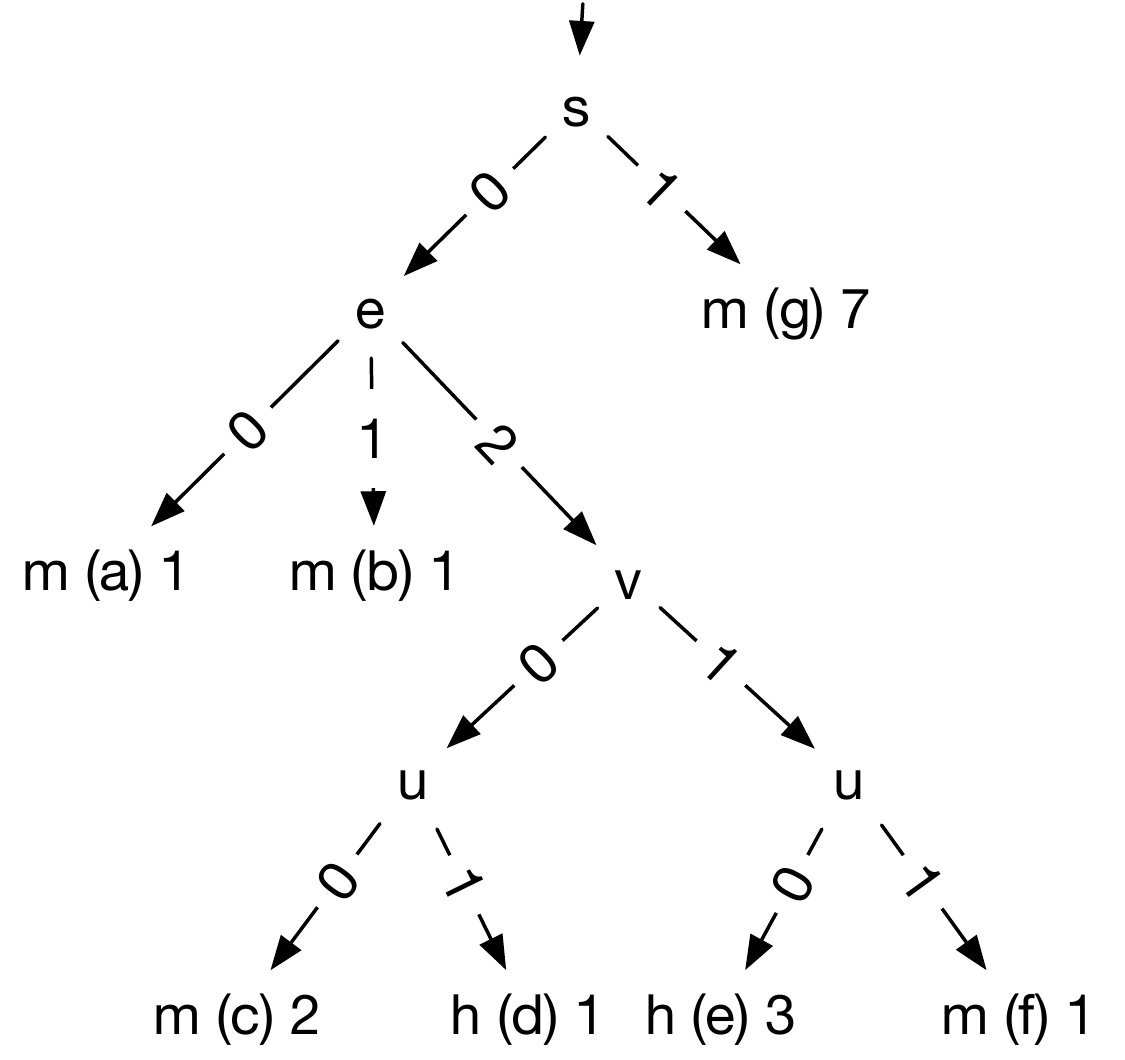}
    \end{minipage}
    \caption{New configurations created from paths (c), (d), (e) in the tree in Figure~\ref{fig:f2} and new decision tree for {\exloc}.}
    \label{fig:f3}
  \end{minipage}
\end{figure*}

We use the C program in Figure~\ref{fig:exprog} to explain {\tool}.
This program has nine configuration options listed on the first line of the figure.
The four options $s,t,u,v$ are boolean-valued, and the other five options, $a,b,c,d,e$, range over the set $\{0,1,2\}$.
The \emph{configuration space} of this program thus has $2^4 \times 3^5 = 3888$ possible configurations.

The code in Figure~\ref{fig:exprog} includes print statements that mark nine locations $L0$--$L8$.
At each location, we list the associated desired interaction.
For example, $L5$ is covered by any configuration in which $s$ is \texttt{true}, $e$ is 2, and either $u$ or $v$ is false.
$L0$ is covered by every configuration (i.e., having the interaction \emph{true}), but $L6$ is \emph{not} covered by every configuration  because the program returns when it reaches $L3$ or $L5$.

Prior interaction inference approaches are not sufficient for this example.
The works of Reisner et. al~\cite{reisner2010using} and iTree~\cite{song:icse12,song:2014aa} only support conjunctions and therefore cannot generate the correct interactions for any locations except $L0$, $L2$, and $L3$.
The iGen tool~\cite{nguyen2016igen}, which supports conjunctions, disjunctions, and a limited form of both conjunctions and disjunctions, also cannot generate the interactions for locations $L6$ and $L8$.

\paragraph*{Initial Configurations} {\tool} first creates a random 1-way covering array~\cite{cohen2003constructing,cohen1996combinatorial} to obtain a set of initial configurations, which contains all possible settings of each individual option.
Figure~\ref{fig:f1} shows the initial configurations and their coverage information for the running example.

\paragraph*{Decision Trees} For each covered location $l$, {\tool} uses a classification algorithm called {\myca}, developed specifically for this work, ({\S}\ref{sec:classification}) to build a decision tree representing the interaction for $l$.
To build the tree for $l$, {\myca} uses two sets of data: the \emph{hit} sets consisting of configurations covering $l$ and the \emph{miss} set consisting of configurations not covering $l$.
For example, for {\exloc}, {\tool} builds the decision tree in Figure~\ref{fig:f1} from the hit sets $\{c_2\}$ and the miss set $\{c_1, c_3\}$.

From the given configurations  {\myca} determines that the coverage of {\exloc} just requires option $s$ being 0 (\texttt{false}).
Thus, the interaction for {\exloc}, represented by the condition of the \emph{hit} path (a) of the tree in Figure~\ref{fig:f1}, is $\bar{s}$.
This interaction is quite different than  $\bar{s} \wedge e\equiv2  \wedge ((u \wedge \bar{v}) \vee (\bar{u} \wedge v))$, the desired interaction for {\exloc}.
However, even with only three initial configurations, the tree is partially correct because configurations having $s$ as true would miss {\exloc} and $s$ being false is part of the requirements for hitting {\exloc}.

\paragraph*{New Configurations} {\tool} now attempts to create new configurations to refine the tree representing the interaction for location $l$.
Observe that if a hit path is precise, then any configuration satisfying its condition would cover $l$ (similarly, any configuration satisfying the condition of a miss path would not cover $l$).
Thus, we can validate a path by generating configurations satisfying its condition and checking their coverage.
Configurations generated from a hit (or miss) path that do not (or do) cover $l$ are \emph{counterexample} configurations, which show the imprecision of the path condition and help build a more precise tree in the next iteration.

In the running example, {\tool} selects the condition $\bar{s}$ of the hit path (a) of the tree shown in Figure~\ref{fig:f1} and generates four new configurations shown in Figure~\ref{fig:f2} with $s=0$ and 1-covering values for the other eight variables.
If path (a) is precise, then these configurations would cover {\exloc}.
However, only configuration $c_7$ covers {\exloc}.
Thus, $c_4,c_5,c_6$, which do not cover {\exloc}, are counterexamples showing that path (a) is imprecise and thus $\bar s$ is not the correct interaction for {\exloc}.

Note that we could also generate new configurations using  path (b), which represents the interaction for \emph{not} covering {\exloc}.
However, {\tool} prefers path (a) because the classifier uses one configuration for path (a) and two for path (b), i.e., the condition $\bar{s}$ for covering $l$ is only supported by one configuration and thus is likely more imprecise.

\paragraph*{Next Iterations} {\tool} now repeats the process of building trees and generating new configurations.
Continuing with our example on finding the interaction for {\exloc}, {\tool} adds $c_7$ to the hit set and $c_4,c_5,c_6$ to the miss set and builds the new tree for {\exloc} in Figure~\ref{fig:f2}.
The combination of the hit paths (d) and (e) gives $e\equiv 2 \wedge (u \vee (\bar{u}\wedge v))$ as the interaction for {\exloc}.
This interaction contains options $e,u,v$, which appear in the desired interaction $\bar{s} \wedge e\equiv2 \wedge ((u \wedge \bar{v}) \vee (\bar{u} \wedge v))$.

To validate the new interaction for {\exloc}, {\tool} generates new configurations from paths (c), (d), (e) of the tree in Figure~\ref{fig:f2}, because they have the fewest number of supporting configurations.
Figure~\ref{fig:f3} shows the nine new configurations.

Note that (c) is a miss path and thus $c_8, c_9, c_{10}$ are \emph{not} counterexamples because they do not hit {\exloc}.
Also, in an actual run, {\tool} would select only one of these three paths and take two additional iterations to obtain these configurations.
For illustration purposes, we combine these iterations and show the generated configurations all together.

In the next iteration, using the new configurations and the previous ones, {\tool} builds the decision tree in Figure~\ref{fig:f3} for {\exloc}.
The interaction obtained from the two hit paths (d) and (e) is $\bar{s} \wedge e\equiv 2 \wedge ((\bar{v} \wedge u) \vee (v \wedge \bar{u}))$, which is equivalent to the desired one and thus would remain unchanged regardless of any additional configurations {\tool} might create.%

Finally, {\tool} stops when it cannot generate new coverage or refine existing trees for several consecutive iterations.
In a postprocessing step, {\tool} combines the hit path conditions of the decision tree for each location $l$ into a logical formula representing the interaction for $l$. %

\paragraph*{Complete Run} {\tool} found the correct interactions for all locations in the running example within eight iterations and under a second. %
The table below shows the number of iterations and configurations used to find the interaction for each location.
For example, the desired interaction for {\exloc} took 58 configurations and is discovered at iteration 4, and the interaction \texttt{true} of L0 was quickly discovered from the initial configurations.

\begin{table}[h!]
  \centering
  \begin{tabular}{l|r|r|r|r|r|r|r|r|r|r }
    & L0 & L1 & L2 & L3 & L4 & L5 & L6 & L7 & L8 \\
    \hline
    Iter. Found   & 1 & 2 & 6 & 1 & 2 & 5 & 3 & 3 & 4 \\
    \# Configs& 3 & 27 & 144 & 15 & 30 & 123 & 50 & 47 & 58
  \end{tabular}
\end{table}

Overall, {\tool} found all of these interactions by analyzing approximately 360 configurations (median over 11 runs) out of 3888 possible ones.
The experiments in {\S}\ref{sec:eval} show that {\tool} analyzes an even smaller fraction of the possible configurations on programs with larger configuration spaces.

\section{Preliminaries}
A configurable software consists of multiple \emph{configuration options}, where each option plays a similar role as a global program variable, but often has a finite domain (e.g., boolean) and does not change during program execution.
A \emph{configuration} is a set of \emph{settings} of the form $x=v$, where $x$ is a configuration option and $v$ is a (valid) value of $x$.

\paragraph{Interactions} An interaction for a location $l$ characterizes of the set of configurations covering $l$.
For example, we see from Figure~\ref{fig:exprog} that any configuration satisfying $u \wedge v$ (i.e., they have the settings $u=1$ and $v=1$) is guaranteed to cover L3.
Although we focus on location coverage, interaction can be associated with more general program behaviors, e.g., we could use an interaction to characterize configurations triggering some undesirable behavior.
To obtain coverage, we typically run the program using a configuration and \emph{a test suite}, which is a set of fixed environment data or options to run the program on, e.g., the test suite for the Unix \texttt{ls} (listing) command might consist of directories to run \texttt{ls} on.
In summary, we define program interactions as:
\begin{definition}
  Given a program $P$, a test suite $T$, and a coverage criterion $X$ (e.g., some location $l$ or behavior $b$), an \emph{interaction} for $X$ is a formula $\alpha$ over the (initial settings of the) configuration options of $P$ such that (a) any configuration satisfying $\alpha$ is guaranteed to cover $X$ under $T$ and (b) $\alpha$ is the logically weakest such formula (i.e., if $\beta$ also describes configurations covering $X$ then $\beta \Rightarrow \alpha$).%
\end{definition}

\paragraph{Decision Trees}
We use a decision tree to represent the interaction for a location $l$.
A decision tree consists of a root, leaves, and internal (non-leaf) nodes.
Each non-leaf node is labeled with a configuration option and has $k$ outgoing edges, which correspond to the $k$ possible values of the option.
Each leaf is labeled with a hit or miss class, which represents the classification of that leaf.
The path from the root to a leaf represents a condition leading to the classification of the leaf.
This path condition is the conjunction of the settings collected along that path.
The union (disjunction) of the hit conditions is the interaction for $l$.
Dually, the disjunction of the miss conditions is the condition for not covering $l$.
The \emph{length} of a path is the number of edges in the path.

For illustration purposes, we annotate each leaf with a label $t\; (a)\; k$, where $t$ is either the (h) hit or (m) miss class, $a$ is the path name (so that we can refer to the path), and $k$ is the number of supporting configurations used to classify this path.
Intuitively, the more supporting configurations a path has, the higher confidence we have about its classification.

For example, the decision tree in Figure~\ref{fig:f3} for location {\exloc} consists of four internal nodes and seven leaves. The tree has five miss and two hit paths, e.g., path (d), which has length 4 and condition $\bar s \wedge e \equiv 2 \wedge \bar v \wedge u$, is classified as a hit due to one configuration hitting {\exloc} ($c_2$ in Figure~\ref{fig:f1}), and (g) is a miss path with condition $s$ because seven configurations satisfying this condition miss {\exloc}.
The interaction for {\exloc} is $\bar{s} \wedge e\equiv 2 \wedge ((\bar{v} \wedge u) \vee (v \wedge \bar{u}))$, the disjunction of the two hit conditions.

\SetKwInOut{Input}{input}
\SetKwInOut{Output}{output}
\SetKwData{hits}{hits}
\SetKwData{misses}{misses}

\section{The {\tool} Algorithm}
\label{sec:algorithm}

\begin{algorithm}[t]
  \small
  \DontPrintSemicolon

  \SetKwData{configs}{configs}
  \SetKwData{cov}{cov}
  \SetKwData{newcov}{new\_cov}
  \SetKwData{true}{true}
  \SetKwData{false}{false}
  \SetKwData{trees}{trees}
  \SetKwData{oldtrees}{old\_trees}
  \SetKwData{paths}{paths}
  \SetKwData{isstable}{is\_stable}
  \SetKwData{interactions}{interactions}
  \SetKwData{countstable}{explore\_iters}
  \SetKwData{maxcountstable}{max\_explore\_iters}
  \SetKwData{needrebuild}{need\_rebuild}
  \SetKwData{exploremode}{explore\_mode}
  \SetKwData{location}{location}

  \SetKwFunction{run}{run}
  \SetKwFunction{mybreak}{break}
  \SetKwFunction{gennewconfigs}{gen\_new\_configs}
  \SetKwFunction{hit}{hit}
  \SetKwFunction{miss}{miss}
  \SetKwFunction{testtree}{test\_tree}
  \SetKwFunction{buildtree}{build\_tree}
  \SetKwFunction{postprocess}{post\_process}
  \SetKwFunction{selectpaths}{select\_ranked\_paths}
  \SetKwFunction{selectpathsrand}{select\_random\_paths}
  \SetKwFunction{onewaycovering}{oneway\_covering\_configs}
  \SetKwFunction{isnull}{is\_null}

  \Input{program $P$; test suite $T$; initial configs $I$ (optional)}
  \Output{a set of interactions of $P$}
  \BlankLine

  $\configs \leftarrow I ~\cup~ \onewaycovering{}$\; \label{line:init}
  $\cov \leftarrow \run(P,T,\configs)$\; \label{line:testsuite0}
  $\trees \leftarrow \emptyset$\;
  \BlankLine

  $\countstable \leftarrow 0$ \;
  \While{$\countstable < \maxcountstable$}{ \label{line:whileloop}
    $\countstable \leftarrow \countstable + 1$\;
    $\exploremode \leftarrow \countstable > 1$\;

    \ForEach {\location $l \in \cov$}{ \label{line:forloop}
      $\hits \leftarrow \hit(\cov,l)$\;
      $\misses \leftarrow \miss(\cov,l)$\;
      \BlankLine
      $\needrebuild \leftarrow \isnull(\trees[l]) \lor \neg{\testtree}(\trees[l], \hits, \misses)$\;
      \If{$\needrebuild \vee \exploremode$}{
      	\If{$\needrebuild$} {
        	$\countstable \leftarrow 0$\;
        	$\trees[l] \leftarrow \buildtree(\hits, \misses)$\;
       	}
       \BlankLine
        $\paths \leftarrow \selectpaths(\trees[l])$\;
        \If{$\exploremode$} {
        	$\paths \leftarrow \paths \cup \selectpathsrand(\trees[l])$\;
        }
    	\BlankLine
        $\configs \leftarrow \gennewconfigs(\paths)$\;  \label{line:gennewconfigs}
        $\cov \leftarrow \cov \cup \run(P,T,\configs)$\;
      }
    }
  }
  $\interactions \leftarrow \postprocess(\trees)$\;
  \KwRet{\interactions}
  \caption{{\tool}'s iterative refinement algorithm}
  \label{alg:igen}
\end{algorithm}

Figure~\ref{alg:igen} shows the {\tool} algorithm, which takes as input a program, a test suite, and an optional set of initial configurations, and returns a set of interactions for locations in the program that were covered.
Initial configurations, e.g., default or factory-installed configurations, if available, are useful starting points because they often give high coverage.

{\tool} starts by creating a set of configurations using a randomly generated 1-covering array and the initial configurations if they are available.
{\tool} then runs the program on \texttt{configs} using the test suite and obtain their coverage.

Next, {\tool} enters a loop that iteratively builds a decision tree for each covered location ({\S}\ref{sec:classification}) and generates new configurations from these trees ({\S}\ref{sec:gen_configs}) in order to refine them.
{\tool} has two modes: \texttt{exploit} and \texttt{explore}. It starts in \texttt{exploit} mode and refines incorrect trees in each iteration.
When {\tool} can no longer refine trees (e.g., it is stuck in some plateau), it switches to \texttt{explore} mode and generates random configurations, hoping that these could help improve the trees (and if so, {\tool} switches back to \texttt{exploit} mode in the next iteration).

For each covered location $l$, {\tool} performs the following steps.
First, we create hit and miss sets consisting of configurations hitting or missing $l$, respectively.
Second, if {\tool} is in \texttt{exploit} mode, we build a decision tree for $l$ from the hit and miss sets of configurations if either $l$ is a new location (a tree for $l$ does not exist) or that the existing tree for $l$ is not correct (the \texttt{test\_tree} function checks if the tree fails to classify some configurations). 
If both of these are \emph{not true} (i.e., the existing tree for $l$ is correct), we continue to the next location.
Otherwise, if {\tool} is in \texttt{explore} mode, we continue to the next step.
Third, we rank and select paths in the tree that are likely incorrect to refine them. If {\tool} is in \texttt{explore} mode, we also select random paths.
Finally, we generate new configurations using the selected paths and obtain their coverage. {\tool} uses these configurations to validate and refine the decision tree for $l$ in the next iteration.

{\tool} repeats these steps until existing trees remain the same and no new trees are generated (i.e., no new coverage) for several iterations.
In the end, {\tool} uses a postprocessing step to extract logical formulae from generated trees to represent program interactions.

\subsection{Selecting Paths and  Generating Configurations}
\label{sec:gen_configs}

\begin{figure}
  \begin{minipage}{0.49\linewidth}
      \centering
    \includegraphics[width=0.9\linewidth]{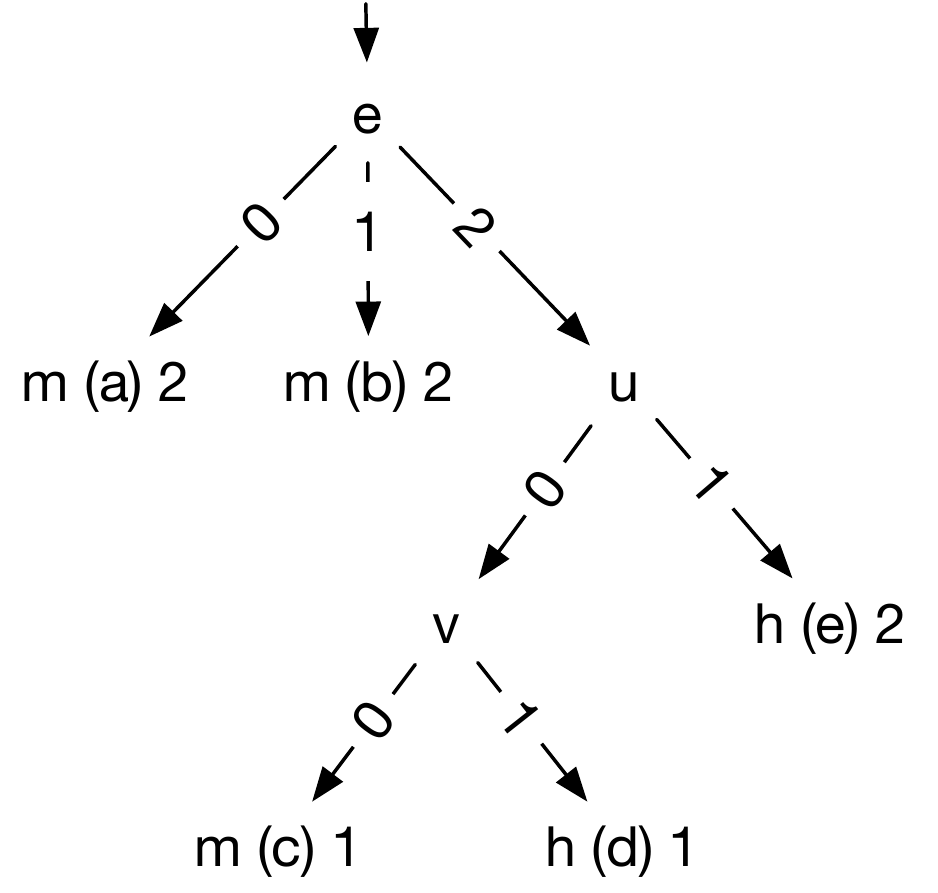}
  \end{minipage}
  \hfill
  \begin{minipage}{0.50\linewidth}
    \centering
    \[
      \small
      \begin{array}{rrrrrrrrr}
      s & t & u & v & a & b & c & d & e  \\
      \midrule
      0 & 0 & 0 & 0 & 0 & 0 & 2 & 2 & 2  \\
      1 & 1 & 0 & 0 & 2 & 1 & 1 & 0 & 2  \\
      1 & 0 & 0 & 0 & 1 & 2 & 0 & 1 & 2  \\
      \end{array}
    \]
  \end{minipage}
  \caption{A decision tree and new configurations created from path (c) of the tree}
  \label{fig:dt4}
\end{figure}

Given a decision tree, {\tool} ranks paths in the tree and generates new configurations from high-ranked ones.
Intuitively, we use configurations generated from a path to validate that path condition, which represents an interaction.
If these configurations do not violate the path condition, we gain confidence in the corresponding interaction.
Otherwise, these configurations are counterexamples that are subsequently used to learn a new tree with more accurate paths. 

\paragraph{Selecting Paths} To select paths to generate new configurations, {\tool} favors those with \emph{fewer} supporting configurations because such paths are likely inaccurate and thus generating counterexample configurations to ``break'' them is likely easier.

If there are multiple paths with a similar number of supporting configurations, we break ties by choosing the \emph{longest} ones.
Paths with few supporting configurations but involving many options are likely more fragile and inaccurate.
If there are multiple paths with a similar length and number of supporting configurations, we pick one arbitrary.

For example, paths (c) and (d) in the tree shown in Figure~\ref{fig:dt4} have the highest rank because they each have just one supporting configuration.
Paths (a), (b), and (e) have two configurations each, but path (e) is longer and thus ranked higher.
The final ranking for this tree is then (c), (d), (e), (a), and (b).

\paragraph{Generating Configurations} From the highest-ranked path, {\tool} generates 1-covering configurations that satisfy the path condition, i.e., these configurations have the same settings as those in the condition of that path.
{\tool} keeps generating new configurations this way for the next highest-ranked paths until it achieves up to a certain number of new configurations (currently configured to generate at least two new configurations).%

Using high-ranked paths to generate configurations is a greedy approach, which might not always give useful configurations that help improve the tree.
Thus, {\tool} also selects random paths during the \texttt{explore} mode, i.e., when a tree remains unchanged in the previous iteration so that lower-ranked paths can also be improved.%

Figure~\ref{fig:dt4} shows one possible set of configurations generated from the highest-ranked path \texttt{c}.
The condition of path \texttt{c} is $e \equiv 2 \wedge u \equiv 0 \wedge v \equiv 0$ and thus all generated configurations have values of $e,u,v$ fixed to $2,0,0$, respectively.

\subsection{Building Decision Trees}
\label{sec:classification}

{\tool} uses a specialized classification algorithm to build decision trees.
While many decision tree classifiers exist (e.g., the popular family of ID3, C4.5, and C5.0 algorithms~\cite{quinlan:2014aa,kuhn:2013aa}), they do not fit our purpose because they employ aggressive pruning strategies to simplify trees and need large dataset to produce accurate results.

\subsubsection{Limitations of C5.0}\label{sec:c5.0limits} Consider an example where we have three options: $s,t$ are bool and $z$ ranges over the values $\{0,1,2,3,4\}$.
Assume we use all $2\times 2 \times 5=20$ configurations as sample data and use the interaction $s \land t \land (1 \le z \land z \le  3)$ to classify these configurations: 3 hits (there are only 3 configurations satisfy this interaction) and 17 misses.

\begin{figure}
  \begin{minipage}{0.44\linewidth}
    \centering
    \includegraphics[width=1.3\linewidth]{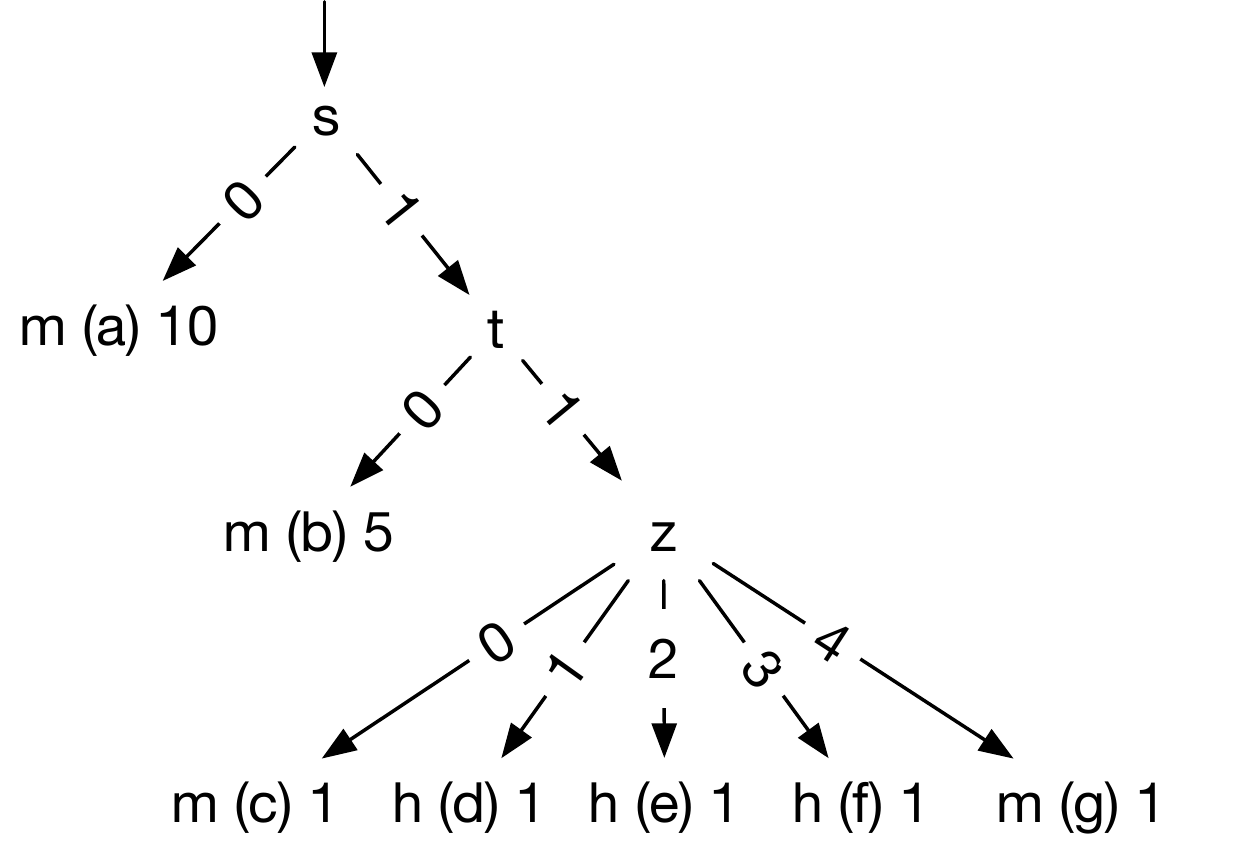}
    \caption*{(a)}
  \end{minipage}
  \hfill
  \begin{minipage}{0.16\linewidth}
    \centering
    \includegraphics[width=0.8\linewidth]{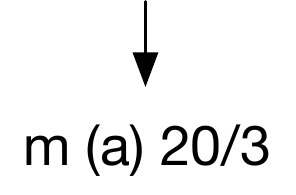}
    \caption*{(b)}
  \end{minipage}
  \hfill
  \begin{minipage}{0.34\linewidth}
    \centering
    \includegraphics[width=1.0\linewidth]{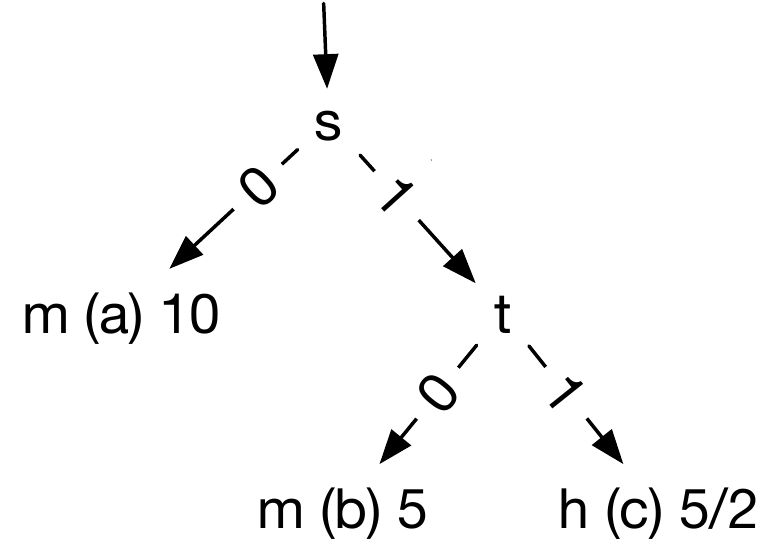}
    \caption*{(c)}
  \end{minipage}
  \caption{Ideal tree (a) and C5.0 trees (b,c)}
  \label{fig:compare}
\end{figure}

The C5.0 algorithm would not be able to create a decision tree, e.g., the one shown in Figure~\ref{fig:compare}a, that perfectly classifies this data set to represent the desired interaction.
For example, the official C5.0 implementation~\cite{c50website} with default settings yields the tree in Figure~\ref{fig:compare}b, which represents the interaction \texttt{False}.
This is because by default, the tool determines that most samples were misses (17/20) and prunes nodes to create a tree reflecting this belief\footnote{The label 20/3 indicates this classification has a total of 20 samples, but 3 of them are incorrect.}.
After tweaking the tool's parameters to avoid pruning\footnote{Using the custom parameters \texttt{-c 100 -m 1 -g}.}, we obtain the tree in Figure~\ref{fig:compare}c that represents the interaction $s \wedge t$, which is more accurate, but is still far from the desired one shown in Figure~\ref{fig:compare}a.
Even with this full set of configurations, we cannot modify C5.0 to obtain the desired interaction, because 
C5.0, like many other ML techniques, requires a \emph{very large} set of sample data to be accurate (leaves with too few samples, e.g., the 3 hit configurations in this example, are given low ``confidence level'' and therefore are pruned).

\subsubsection{The {\myca} algorithm}
\label{sec:myca}

We develop {\myca}, a ``simplified'' version of C5.0 for interaction learning.
Similarly to C5.0, {\myca} builds a decision tree to split a training sample (e.g., hit and miss configurations) based on the feature (e.g., configuration options) that provides the highest information gain.
Each subsample is then split again using a different feature, and the process repeats until meeting some stopping criteria.

Classification algorithms including ID3, C4.5, C5.0, CART are designed around the concept of pruning, i.e., ``remove parts of the tree that do not contribute to classification accuracy on unseen cases, producing something less complex and thus more comprehensible''~\cite{quinlan:2014aa}.
But pruning leads to inaccuracy as shown in {\S}\ref{sec:c5.0limits}.
Thus, {\myca} avoids pruning to achieve a 100\% accuracy on the training sample, i.e., every sample configuration is correctly classified.

Other than pruning, the two algorithms have several main differences.
First, we use two classification categories (hit and miss) and features (configuration options) with finite domains, e.g., boolean or over a finite set of values.
Our training samples do not contain unknown values (C5.0 allows some values in the training data to be omitted).
The sample data also does not contain noise, e.g., if $c$ is an interaction for a location, then any configuration satisfies $c$ will guarantee to hit $c$.
We give similar weights to samples and similar costs for misclassifications (C5.0 allows different cost assignments to misclassification).
Finally, we perform splitting until we can no longer split subsamples while C5.0 uses heuristics to decide when to stop splitting and prune the rest.

Using the set of 20 configurations in the example in {\S}\ref{sec:c5.0limits}, {\myca} generates the tree in Figure~\ref{fig:compare}a, which represents the desired interaction.  In fact, {\myca} can generate the same tree using just 14 configurations. %
However, by requiring exact, instead of more generalized, trees, {\myca} is prone to ``overfitting'', i.e., generating trees that are correct for the sample data but might not in general.
{\tool}'s iterative refinement phase is specifically designed to mitigate this problem, i.e., by generating counterexample configurations to gradually correct overfitting mistakes.
In {\S}\ref{sec:eval}, we show that the integration of {\myca} and iterative refinement helps {\tool} scale to programs with very large configuration spaces and learn trees representing accurate interactions using  small sets of configurations.

\section{Subject Programs}
\label{sec:subjects}

{\tool} is implemented in C++ and uses the Z3 SMT solver~\cite{z3ms} to encode and simplify interactions.
We also use Z3 to analyze interactions as described in {\S}\ref{sec:eval} (e.g., checking that interactions are equivalent to ground truth).

\subsection{Subject Programs}
\label{sec:subs}

\begin{table}[t]
\centering
\caption{Subject programs}
\label{tab:stats}
\begin{tabular}{llrrrr}
  \textbf{prog} & \textbf{lang} & \textbf{ver} & \textbf{loc} & \textbf{opts}  & \textbf{cspace}  \\
  \midrule
  id    & C & 8.32 & 342  & 10 & 1024       \\
  uname & C & 8.32 & 282  & 11 & 2048       \\
  cat   & C & 8.32 & 484  & 12 & 4096       \\
  mv    & C & 8.32 & 378  & 11 & 5120       \\
  ln    & C & 8.32 & 521  & 12 & 10 240     \\
  date  & C & 8.32 & 501  &  7 & 17 280     \\
  join  & C & 8.32 & 895  & 12 & 18 432     \\
  sort  & C & 8.32 & 3366 & 22 & 6 291 456  \\
  ls    & C & 8.32 & 3972 & 47 & $3.5 \times 10^{14}$ \\
  \midrule
  grin  & Python & 1.2.1 & 628    & 22 & 4 194 304     \\
  pylint& Python & 1.9.5 & 15 493 & 28 & $2.9 \times 10^{11}$  \\
  unison  & Ocaml & 2.51.2 & 30 074 & 27 & $2.0 \times 10^8$     \\
  bibtex2html  & Ocaml & 1.99 & 9258 & 33 & $1.3 \times 10^{10}$     \\
  cloc  & Perl & 1.86  & 12 427 & 23 &  16 777 216   \\
  ack   & Perl & 3.4.0 & 3244  & 28 &  $5.4 \times 10^8$    \\
  \midrule
  vsftpd  & C & 2.0.7 &  10 482 & 30 & $2.1 \times 10^9$  \\
  ngircd  & C & 0.12.0 & 13 601 & 13 & 294 912  \\
  \bottomrule
\end{tabular}
\end{table}

\begin{table*}
  \centering
  \caption{Results. Column \texttt{min cspace} lists the results for the experiment in {\S}\ref{sec:analysis}}
  \label{tab:results}
  \begin{tabular}{l|rlrl|rlrl|rlrlrlrlrl|rlrl||c}
\multicolumn{1}{c|}{}&\multicolumn{4}{c|}{}&\multicolumn{4}{c|}{\textbf{time(s)}}&\multicolumn{10}{c|}{\textbf{interaction types}}&\multicolumn{4}{c||}{\textbf{inter. lengths}}&\textbf{min}\\
    \textbf{prog} &
    \multicolumn{2}{c}{\textbf{configs}}&
    \multicolumn{2}{c|}{\textbf{cov}}&
    \multicolumn{2}{c}{\textbf{search}}&
    \multicolumn{2}{c|}{\textbf{total}}&
    \multicolumn{2}{c}{\textbf{single}}&
    \multicolumn{2}{c}{\textbf{conj}}&
    \multicolumn{2}{c}{\textbf{disj}}&
    \multicolumn{2}{c}{\textbf{mix}}&
    \multicolumn{2}{c|}{\textbf{total}}&
    \multicolumn{2}{c}{\textbf{max}}&
    \multicolumn{2}{c||}{\textbf{median}}&
    \textbf{cspace}\\
    \midrule
    id    & 609 & {\tiny 277} & 150 & {\tiny 1} & 0 & {\tiny 0} & 0 & {\tiny 0} & 3 & {\tiny 0} & 17 & {\tiny 3} & 1 & {\tiny 1} & 11 & {\tiny 5} & 32 & {\tiny 8} & 10 & {\tiny 2} & 5 & {\tiny 2} & 10  \\
    uname & 189 & {\tiny 359} & 98 & {\tiny 1} & 0 & {\tiny 0} & 0 & {\tiny 0} & 11 & {\tiny 0} & 4 & {\tiny 1} & 0 & {\tiny 1} & 8 & {\tiny 2} & 23 & {\tiny 3} & 6 & {\tiny 4} & 1 & {\tiny 0} & 4  \\
    cat   & 1660 & {\tiny 109} & 205 & {\tiny 0} & 0 & {\tiny 0} & 1 & {\tiny 0} & 12 & {\tiny 0} & 7 & {\tiny 0} & 1 & {\tiny 0} & 7 & {\tiny 0} & 27 & {\tiny 0} & 12 & {\tiny 0} & 1 & {\tiny 0} & 6  \\
    mv    & 4532 & {\tiny 61} & 167 & {\tiny 0} & 5 & {\tiny 1} & 6 & {\tiny 1} & 9 & {\tiny 0} & 3 & {\tiny 0} & 3 & {\tiny 0} & 6 & {\tiny 0} & 21 & {\tiny 0} & 11 & {\tiny 0} & 1 & {\tiny 0} & 4  \\
    ln    & 2143 & {\tiny 114} & 171 & {\tiny 0} & 0 & {\tiny 0} & 2 & {\tiny 0} & 10 & {\tiny 0} & 7 & {\tiny 0} & 2 & {\tiny 0} & 5 & {\tiny 0} & 24 & {\tiny 0} & 8 & {\tiny 0} & 1 & {\tiny 0} & 5  \\
    date  & 12050 & {\tiny 741} & 125 & {\tiny 0} & 4 & {\tiny 1} & 7 & {\tiny 1} & 2 & {\tiny 0} & 3 & {\tiny 0} & 2 & {\tiny 0} & 10 & {\tiny 0} & 17 & {\tiny 0} & 6 & {\tiny 0} & 6 & {\tiny 0} & 7  \\
    join  & 4001 & {\tiny 797} & 365 & {\tiny 0} & 1 & {\tiny 1} & 4 & {\tiny 1} & 4 & {\tiny 0} & 17 & {\tiny 1} & 2 & {\tiny 0} & 10 & {\tiny 1} & 33 & {\tiny 0} & 12 & {\tiny 0} & 12 & {\tiny 0} & 6  \\
    sort  & 141935 & {\tiny 23903} & 1085 & {\tiny 0} & 744 & {\tiny 310} & 1069 & {\tiny 345} & 12 & {\tiny 0} & 5 & {\tiny 0} & 2 & {\tiny 0} & 132 & {\tiny 0} & 151 & {\tiny 0} & 22 & {\tiny 0} & 16 & {\tiny 0} & 18  \\
    ls    & 112566 & {\tiny 26356} & 1289 & {\tiny 0} & 31 & {\tiny 5} & 579 & {\tiny 66} & 46 & {\tiny 0} & 40 & {\tiny 0} & 1 & {\tiny 0} & 106 & {\tiny 0} & 193 & {\tiny 0} & 47 & {\tiny 0} & 4 & {\tiny 0} & 14  \\
    \midrule
    grin    & 1828 & {\tiny 132} & 332 & {\tiny 0} & 0 & {\tiny 1} & 103 & {\tiny 7} & 3 & {\tiny 0} & 5 & {\tiny 0} & 0 & {\tiny 0} & 9 & {\tiny 0} & 17 & {\tiny 0} & 7 & {\tiny 0} & 5 & {\tiny 0} & 6  \\
    pylint  & 6850 & {\tiny 27} & 10757 & {\tiny 0} & 26 & {\tiny 1} & 3868 & {\tiny 14} & 1 & {\tiny 0} & 12 & {\tiny 0} & 2 & {\tiny 0} & 8 & {\tiny 0} & 23 & {\tiny 0} & 7 & {\tiny 0} & 5 & {\tiny 0} & 6  \\
    unison  & 9690 & {\tiny 790} & 3565 & {\tiny 0} & 5 & {\tiny 0} & 252 & {\tiny 12} & 3 & {\tiny 0} & 39 & {\tiny 0} & 1 & {\tiny 0} & 19 & {\tiny 0} & 62 & {\tiny 0} & 10 & {\tiny 0} & 7 & {\tiny 0} & 7  \\
    bibtex2html & 38317 & {\tiny 2311} & 1437 & {\tiny 0} & 26 & {\tiny 4} & 5195 & {\tiny 201} & 35 & {\tiny 0} & 11 & {\tiny 0} & 1 & {\tiny 0} & 113 & {\tiny 0} & 160 & {\tiny 0} & 33 & {\tiny 0} & 1 & {\tiny 0} & 10  \\
    cloc & 12284 & {\tiny 931} & 1147 & {\tiny 0} & 612 & {\tiny 1} & 9494 & {\tiny 530} & 3 & {\tiny 0} & 35 & {\tiny 0} & 1 & {\tiny 0} & 20 & {\tiny 0} & 59 & {\tiny 0} & 9 & {\tiny 0} & 5 & {\tiny 0} & 9  \\
    ack & 65553 & {\tiny 6708} & 1212 & {\tiny 2} & 13 & {\tiny 5} & 10214 & {\tiny 603} & 2 & {\tiny 0} & 10 & {\tiny 1} & 1 & {\tiny 0} & 59 & {\tiny 4} & 72 & {\tiny 5} & 28 & {\tiny 1} & 22 & {\tiny 1} & 14  \\
    \midrule
    vsftpd & 10920 & {\tiny 614} & 2549 & {\tiny 0} & 2 & {\tiny 1} & 6 & {\tiny 1} & 5 & {\tiny 0} & 36 & {\tiny 0} & 2 & {\tiny 0} & 9 & {\tiny 0} & 52 & {\tiny 0} & 7 & {\tiny 0} & 5 & {\tiny 0} & 5  \\
    ngircd & 28711 & {\tiny 1335} & 3090 & {\tiny 0} & 18 & {\tiny 1} & 148 & {\tiny 4} & 4 & {\tiny 0} & 8 & {\tiny 0} & 2 & {\tiny 0} & 51 & {\tiny 0} & 65 & {\tiny 0} & 11 & {\tiny 0} & 4 & {\tiny 0} & 5  \\
    \bottomrule
  \end{tabular}
\end{table*}

To evaluate {\tool}, we used the subject programs listed in Table~\ref{tab:stats}.
For each program, we list its name, language, version, and lines of code as measured by \texttt{SLOCCount}~\cite{sloccount}.
We also report the number of configuration options (\texttt{opts}) and the configuration spaces (\texttt{cspace}).

These programs and their setups ({\S}\ref{sec:setup}) are collected from iGen.
We include all programs that we can reproduce the iGen's setup and omit those that we cannot (e.g., the runscripts and tests are not available for the Haskell and Apache \texttt{httpd} used in iGen).
In total, we have 17 programs spanning 4 languages (C, Python, Perl, and Ocaml).

The first group of programs comes from the widely used GNU \texttt{coreutils}~\cite{coreutils}.
These programs are configured via command-line options.
We used a subset of \texttt{coreutils} with relatively large configuration spaces (at least 1024 configurations each).
The second group contains an assortment of programs to demonstrate {\tool}'s wide applicability.
Briefly: \texttt{grin} and \texttt{ack} are grep-like programs;
\texttt{pylint} is a static checker for Python;
\texttt{unison} is a file synchronizer;
\texttt{bibtex2html} converts BibTeX files to HTML; 
and \texttt{cloc} is a lines of code counter.
These programs are written in Python, Ocaml, and Perl and have the configuration space size ranging from four million to $10^{11}$.
The third group contains \texttt{vsftpd}, a secure FTP server, and \texttt{ngircd}, an IRC daemon.
These programs were also studied by~\cite{reisner2010using}, who uses the Otter symbolic execution tool to exhaustively compute all possible program executions under all possible settings.
Rather than using a test suite, we ran {\tool} on these programs in a special mode in which we used Otter's outputs as an oracle that maps configurations to covered lines.

\subsection{Setup}
\label{sec:setup}
We selected configuration options in a variety of ways.
For \texttt{coreutils} programs, we used all options, most of which are boolean-valued, but nine can take on a wider but finite range of values, all of which we included, e.g., all possible string formats the program \texttt{date} accepts.
We omit options that range over an unbounded set of values.
For the assorted programs in the second group, we used the options that we could get working correctly and ignore those that can take arbitrary values, e.g., \texttt{pylint} options that take a regexp or Python expression as input.
For \texttt{vsftpd} and \texttt{ngircd} we used the same options as in iGen.

We manually created tests for \texttt{coreutils} to cover common usage.
For example, for \texttt{cat}, we wrote a test that read data from a normal text file.
For \texttt{ls}, we let it list the files from a directory containing some files, some subdirectories, and some symbolic links.

Finally, we obtained line coverage using \texttt{gcov}~\cite{gcov} for C,  \texttt{coverage}~\cite{python-cov} for Python, \texttt{Devel::Cover}~\cite{mdevel} for Perl, and expression coverage using \texttt{Bisect}~\cite{Bisect} for OCaml.
We used a custom runner to get the coverage for \texttt{vsftpd} and \texttt{ngircd} using Otter's result as explained in {\S}\ref{sec:subs}.

Our experiments were performed on a 64-core AMD CPU 2.9GHz Linux system with 64 GB of RAM.
{\tool} and all experimental data are available at~\cite{toolwebsite}.

\section{Evaluation}
\label{sec:eval}
To evaluate {\tool} we consider four research questions:
can {\tool} learn accurate program interactions (\emph{R1-Accuracy})? 
how does it perform and scale to programs with large configuration spaces (\emph{R2-Performance})? 
what can we learn from the discovered interactions (\emph{R3-Analysis})? and
how does {\tool} compare to iGen (\emph{R4-Comparing to iGen})?

Table~\ref{tab:results} summarizes the results of running {\tool} on the benchmark programs ({\S}\ref{sec:subjects}), taking median across 11 runs and their variance as the semi-interquartile (SIQR) range~\cite{siqr}.
For each program, columns \texttt{configs} and \texttt{cov} report the number of configurations generated by \tool\ and the number of locations covered by these configurations, respectively.
The next two columns report the running time of {\tool} (\texttt{search} is the \texttt{total} time minus the time spent running programs to obtain coverage).
The next five columns report the number of distinct interactions inferred by {\tool}.
Column \texttt{single} shows the number of interactions that are \texttt{true}, \texttt{false}, or contain only one option, e.g., $\bar s$.
Columns \texttt{conj}, \texttt{disj}, \texttt{mix}, \texttt{total} show the number of pure conjunction, pure disjunction, mixed (arbitrary form), and all of these interactions, respectively.
The low SIQR values on the discovered coverage and interactions indicate that {\tool}, despite being non-deterministic\footnote{{\tool} has several sources of randomness: the initial one-way covering array, the selection of paths used for generating new configurations, the selection of option values in those new configurations, and the creation of the decision tree by the classification algorithm.}, produces relatively stable results across 11 runs.
The next two columns list the max and median interaction lengths, which are further discussed in {\S}\ref{sec:analysis}.
Column {\texttt{min cspace}} lists the results for the experiment discussed in {\S}\ref{sec:analysis}.

\subsection{R1-Accuracy}
\label{sec:correctness}

To measure the accuracy of inferred interactions, we evaluated whether {\tool} produces the same results with its iterative algorithm as it could produce if it used all configurations (i.e., the results {\tool} inferred using \emph{all} configurations are ``ground truths'', representing the real interactions).
To do this comparison, we use all \texttt{coreutils} programs (except \texttt{ls}), \texttt{grin}, and \texttt{ngircd} because we can exhaustively enumerate all configurations for these programs.

\begin{table}[h]
  \centering
  \caption{Comparing {\tool}}
  \label{tab:cmp}
    \centering
    \begin{tabular}{l|crr||crr}
      \multicolumn{1}{c}{}&\multicolumn{3}{c}{\textbf{(a) vs. exhaustive}}&\multicolumn{3}{c}{\textbf{(b) vs. iGen ({\S}\ref{sec:cmp_igen})}}\\      
      \midrule
      &\textbf{cov}&\multicolumn{2}{c||}{\textbf{interactions}}&&\multicolumn{2}{c}{\textbf{mixed}}\\
      prog&$\delta$&\textbf{exact}& \textbf{total}&\textbf{pure}&\textbf{ok}&\textbf{fail}\\
      \midrule
        id    & 0 & 32 & 32 & 21 & 2 & 9  \\
        uname & -1 & 22 & 27 & 17 & 7 & 3  \\
        cat   & 0 & 27 & 27 & 20 & 6 & 1  \\
        mv    & 0 & 21 & 21 & 15 & 2 & 4  \\
        ln    & 0 & 24 & 25 & 20 & 3 & 2  \\
        date  & 0 & 17 & 17 & 7 & 0 & 10  \\
        join  & 0 & 33 & 33 & 23 & 3 & 7  \\
        sort  & 0 & 148 & 151 & 19 & 10 & 122  \\
        \midrule
        grin    & 0 & 17 & 17 & 8 & 9 & 0  \\
        ngircd & 0 & 64 & 65 & 14 & 4 & 47  \\
      \bottomrule
    \end{tabular}
\end{table}

Table~\ref{tab:cmp}a shows the comparison results.
Column \texttt{$\delta$ cov} compares the locations discovered by {\tool} and by exhaustive runs (0 means no difference, $-k$ means {\tool} found $k$ fewer locations).
The next two columns show interactions found by {\tool} (\texttt{exact}) that exactly match the interactions discovered by exhaustive runs (\texttt{total}).

Overall, {\tool} generates highly accurate results comparing to ground truth, while using only a small part of the configuration space as shown in Table~\ref{tab:results} and further described in {\S}\ref{sec:performance}.
For \texttt{uname}, {\tool} misses location \texttt{uname.c:278}, which is guarded by a long conjunction of 11 options of \texttt{uname} (thus the chance of hitting it is 1/2048 configurations). 
Also, for 8/11 times, {\tool} infers inaccurately \texttt{uname.c:202}, which is a long disjunction of 11 options.
For \texttt{ln}, {\tool} was not able to compute the exact interaction for location \texttt{ln.c:495} in all runs.
Manual investigation shows that the interaction of this location is a long disjunction consisting of all 12 run-time options and thus is misidentified by {\tool} as \texttt{true}.
For \texttt{sort}, three locations \texttt{sort.c:3212, sort.c:3492, sort.c:3497} are non-deterministic (running the program on the same configuration might not always hit or miss these locations) and thus produce inaccurate interactions.

\subsection{R2-Performance}
\label{sec:performance}

Table~\ref{tab:results} shows that for programs with large configuration spaces, {\tool} runs longer because it has to analyze more configurations, and the run time is dominated by running the programs on these configurations ($\texttt{total} - \texttt{search}$).
In general, {\tool} scales well to large programs because it only explores a small portion of the configuration space (shown in Table~\ref{tab:stats}).
For small programs (e.g., \texttt{id, uname, cat}), {\tool} analyzes approximately half of the configuration space.
However, for larger programs (e.g., \texttt{sort, ls, pylint, bibtex2html}), {\tool} shows its benefits as the number of configurations analyzed is not directly proportional to the configuration space size. For example, \texttt{ls} has eight more orders of magnitude compared to \texttt{sort}, but the number of explored configurations  is about the same.
Note that \texttt{cloc} and \texttt{ack}'s long run times are due to them being written in Perl, which runs much slower than other languages such as C (and even Python on our machine).

\begin{figure*}
  \tikzset{every picture/.style=semithick}
    \begin{minipage}[b]{0.49\linewidth}
 \begin{tikzpicture}[scale=0.85]
 \begin{axis}[
   grid=both,
   ymin=0.0,
   ymax=1.02,
   xmin=0.0,
   xmax=1.02,
   xlabel={configurations (normalized)},
   ylabel={exact interactions (normalized)},
   legend style={at={(1.3,0.5)},anchor=east,
     nodes={scale=0.9, transform shape}},
   legend columns=1,
 ]

 \addplot[mark=none, color=blue        ] coordinates { (0.0105960264900662,0.09375) (0.0410596026490066,0.25) (0.111258278145695,0.34375) (0.158940397350993,0.59375) (0.196026490066225,0.625) (0.231788079470199,0.875) (0.258278145695364,0.9375) (0.456953642384106,0.96875) (0.462251655629139,0.96875) (0.650331125827815,1) (0.654304635761589,1) (0.835761589403973,1) (1,1)
  };
 \addplot[mark=none, color=blue, dashed] coordinates { (0.0105960264900662,0.1875) (0.0410596026490066,0.34375) (0.111258278145695,0.375) (0.158940397350993,0.40625) (0.196026490066225,0.4375) (0.231788079470199,0.59375) (0.258278145695364,0.59375) (0.456953642384106,0.59375) (0.462251655629139,0.59375) (0.650331125827815,0.625) (0.654304635761589,0.625) (0.835761589403973,0.6875) (1,0.84375)
  };

 \addplot[mark=none, color=green        ] coordinates { (0.0145310435931308,0.111111111111111) (0.0812417437252312,0.592592592592593) (0.103698811096433,0.62962962962963) (0.112945838837517,0.740740740740741) (0.132760898282695,0.740740740740741) (0.157199471598415,0.777777777777778) (0.196829590488771,1) (0.21003963011889,1) (0.367239101717305,1) (0.525099075297226,1) (0.682298546895641,1) (0.844121532364597,1) (1,1)
  };
 \addplot[mark=none, color=green, dashed] coordinates { (0.0145310435931308,0.518518518518518) (0.0812417437252312,0.592592592592593) (0.103698811096433,0.62962962962963) (0.112945838837517,0.62962962962963) (0.132760898282695,0.703703703703704) (0.157199471598415,0.740740740740741) (0.196829590488771,0.740740740740741) (0.21003963011889,0.740740740740741) (0.367239101717305,0.777777777777778) (0.525099075297226,0.777777777777778) (0.682298546895641,0.814814814814815) (0.844121532364597,0.851851851851852) (1,0.851851851851852)
  };

 \addplot[mark=none, color=red        ] coordinates { (0.0139676996944566,0.111111111111111) (0.08642514185945,0.666666666666667) (0.210388476647752,0.814814814814815) (0.236577913574858,1) (0.389786119598429,1) (0.544303797468354,1) (0.701003928415539,1) (0.855958096900917,1) (1,1)
  };
 \addplot[mark=none, color=red, dashed] coordinates { (0.0139676996944566,0.518518518518518) (0.08642514185945,0.62962962962963) (0.210388476647752,0.777777777777778) (0.236577913574858,0.777777777777778) (0.389786119598429,0.814814814814815) (0.544303797468354,0.851851851851852) (0.701003928415539,0.851851851851852) (0.855958096900917,0.888888888888889) (1,0.851851851851852)
  };

 \addplot[mark=none, color=black        ] coordinates { (0.0282836940218556,0.428571428571429) (0.0914934647525177,0.857142857142857) (0.0940647096635955,0.857142857142857) (0.0970644953931862,0.857142857142857) (0.0994214698950075,0.857142857142857) (0.102421255624598,0.857142857142857) (0.104992500535676,0.857142857142857) (0.107349475037497,0.857142857142857) (0.109706449539319,0.857142857142857) (0.174630383544033,0.857142857142857) (0.176344546818084,0.857142857142857) (0.179558602956932,0.857142857142857) (0.181272766230984,0.857142857142857) (0.183844011142061,0.857142857142857) (0.186415256053139,0.857142857142857) (0.188986500964217,0.857142857142857) (0.191557745875295,0.857142857142857) (0.193914720377116,0.857142857142857) (0.19670023569745,0.904761904761905) (0.201414184701093,0.904761904761905) (0.203342618384401,0.904761904761905) (0.205271052067709,0.904761904761905) (0.2082708377973,0.904761904761905) (0.211056353117634,0.904761904761905) (0.212770516391686,0.904761904761905) (0.274480394257553,0.904761904761905) (0.277051639168631,0.904761904761905) (0.278765802442683,0.904761904761905) (0.28197985858153,0.904761904761905) (0.284122562674095,0.904761904761905) (0.340904221127062,0.904761904761905) (0.345403899721448,0.952380952380952) (0.350974930362117,0.952380952380952) (0.353974716091708,0.952380952380952) (0.35568887936576,0.952380952380952) (0.357831583458324,0.952380952380952) (0.359545746732376,0.952380952380952) (0.361688450824941,0.952380952380952) (0.418255838868652,0.952380952380952) (0.420612813370474,0.952380952380952) (0.422755517463038,0.952380952380952) (0.423826869509321,0.952380952380952) (0.479322905506749,0.952380952380952) (0.481251339190058,0.952380952380952) (0.482108420827084,0.952380952380952) (0.535247482322691,0.952380952380952) (0.592243411184915,0.952380952380952) (0.593743304049711,0.952380952380952) (0.644096850224984,0.952380952380952) (0.692736233126205,0.952380952380952) (0.74116134561817,0.952380952380952) (0.784658238697236,0.952380952380952) (0.786158131562031,0.952380952380952) (0.78808656524534,0.952380952380952) (0.831154917505892,0.952380952380952) (0.832440539961431,0.952380952380952) (0.870794943218341,0.952380952380952) (0.908077994428969,0.952380952380952) (0.939790014998929,0.952380952380952) (0.970859224341118,0.952380952380952) (1,0.952380952380952)
  };
 \addplot[mark=none, color=black, dashed] coordinates { (0.0282836940218556,0.666666666666667) (0.0914934647525177,0.904761904761905) (0.0940647096635955,0.904761904761905) (0.0970644953931862,0.904761904761905) (0.0994214698950075,0.904761904761905) (0.102421255624598,0.904761904761905) (0.104992500535676,0.904761904761905) (0.107349475037497,0.904761904761905) (0.109706449539319,0.904761904761905) (0.174630383544033,0.904761904761905) (0.176344546818084,0.904761904761905) (0.179558602956932,0.904761904761905) (0.181272766230984,0.904761904761905) (0.183844011142061,0.904761904761905) (0.186415256053139,0.904761904761905) (0.188986500964217,0.904761904761905) (0.191557745875295,0.904761904761905) (0.193914720377116,0.904761904761905) (0.19670023569745,0.904761904761905) (0.201414184701093,0.904761904761905) (0.203342618384401,0.904761904761905) (0.205271052067709,0.904761904761905) (0.2082708377973,0.904761904761905) (0.211056353117634,0.904761904761905) (0.212770516391686,0.904761904761905) (0.274480394257553,0.904761904761905) (0.277051639168631,0.904761904761905) (0.278765802442683,0.904761904761905) (0.28197985858153,0.904761904761905) (0.284122562674095,0.904761904761905) (0.340904221127062,0.904761904761905) (0.345403899721448,0.904761904761905) (0.350974930362117,0.904761904761905) (0.353974716091708,0.904761904761905) (0.35568887936576,0.904761904761905) (0.357831583458324,0.904761904761905) (0.359545746732376,0.904761904761905) (0.361688450824941,0.904761904761905) (0.418255838868652,0.904761904761905) (0.420612813370474,0.904761904761905) (0.422755517463038,0.904761904761905) (0.423826869509321,0.904761904761905) (0.479322905506749,0.904761904761905) (0.481251339190058,0.904761904761905) (0.482108420827084,0.904761904761905) (0.535247482322691,0.904761904761905) (0.592243411184915,0.904761904761905) (0.593743304049711,0.904761904761905) (0.644096850224984,0.904761904761905) (0.692736233126205,0.904761904761905) (0.74116134561817,0.904761904761905) (0.784658238697236,0.904761904761905) (0.786158131562031,0.904761904761905) (0.78808656524534,0.904761904761905) (0.831154917505892,0.904761904761905) (0.832440539961431,0.904761904761905) (0.870794943218341,0.904761904761905) (0.908077994428969,0.904761904761905) (0.939790014998929,0.904761904761905) (0.970859224341118,0.904761904761905) (1,0.904761904761905)
  };

 \addplot[mark=none, color=brown        ] coordinates { (0.0721500721500721,0.4) (0.25012025012025,0.96) (0.451178451178451,0.96) (0.644059644059644,0.96) (0.828763828763829,0.96) (1,0.96)
  };
 \addplot[mark=none, color=brown, dashed] coordinates { (0.0721500721500721,0.84) (0.25012025012025,0.88) (0.451178451178451,0.88) (0.644059644059644,0.92) (0.828763828763829,0.92) (1,0.92)
  };

 \legend{id, \emph{id}, uname, \emph{uname}, cat, \emph{cat}, mv, \emph{mv}, ln, \emph{ln}}
 \end{axis}
\end{tikzpicture}
\end{minipage}
\hfill
\begin{minipage}[b]{0.49\linewidth}
 \begin{tikzpicture}[scale=0.85]
 \begin{axis}[
   grid=both,
   ymin=0.0,
   ymax=1.02,
   xmin=0.0,
   xmax=1.02,
   xlabel={configurations (normalized)},
   ylabel={exact interactions (normalized)},
   legend style={at={(1.3,0.5)},anchor=east,
     nodes={scale=0.9, transform shape}},
   legend columns=1,
 ]

 \addplot[mark=none, color=blue        ] coordinates { (0.11032741205155,0.411764705882353) (0.147161267850923,0.588235294117647) (0.186955764541971,0.588235294117647) (0.190613026819923,0.588235294117647) (0.192006269592476,0.588235294117647) (0.19339951236503,0.647058823529412) (0.194096133751306,0.647058823529412) (0.194792755137583,0.647058823529412) (0.199582027168234,0.647058823529412) (0.200975269940787,0.647058823529412) (0.371473354231975,0.705882352941177) (0.386015325670498,0.764705882352941) (0.388192267502612,0.764705882352941) (0.389585510275165,0.764705882352941) (0.390978753047719,0.764705882352941) (0.392371995820272,0.764705882352941) (0.393765238592825,0.764705882352941) (0.395158481365378,0.764705882352941) (0.3967258794845,0.823529411764706) (0.397422500870777,0.823529411764706) (0.398119122257053,0.823529411764706) (0.39881574364333,0.823529411764706) (0.399512365029606,0.823529411764706) (0.404040404040404,0.823529411764706) (0.537617554858934,1) (0.539097875304772,1) (0.653082549634274,1) (0.765412748171369,1) (0.860066179031696,1) (0.935301288749565,1) (1,1)
  };
 \addplot[mark=none, color=blue, dashed] coordinates { (0.11032741205155,0.588235294117647) (0.147161267850923,0.647058823529412) (0.186955764541971,0.647058823529412) (0.190613026819923,0.647058823529412) (0.192006269592476,0.647058823529412) (0.19339951236503,0.647058823529412) (0.194096133751306,0.647058823529412) (0.194792755137583,0.647058823529412) (0.199582027168234,0.647058823529412) (0.200975269940787,0.647058823529412) (0.371473354231975,0.647058823529412) (0.386015325670498,0.647058823529412) (0.388192267502612,0.647058823529412) (0.389585510275165,0.647058823529412) (0.390978753047719,0.647058823529412) (0.392371995820272,0.647058823529412) (0.393765238592825,0.647058823529412) (0.395158481365378,0.647058823529412) (0.3967258794845,0.647058823529412) (0.397422500870777,0.647058823529412) (0.398119122257053,0.647058823529412) (0.39881574364333,0.647058823529412) (0.399512365029606,0.647058823529412) (0.404040404040404,0.647058823529412) (0.537617554858934,0.647058823529412) (0.539097875304772,0.647058823529412) (0.653082549634274,0.647058823529412) (0.765412748171369,0.647058823529412) (0.860066179031696,0.647058823529412) (0.935301288749565,0.647058823529412) (1,0.647058823529412)
  };

 \addplot[mark=none, color=green        ] coordinates { (0.022583559168925,0.272727272727273) (0.0726287262872629,0.909090909090909) (0.0789521228545619,0.909090909090909) (0.0829268292682927,0.909090909090909) (0.0861788617886179,0.909090909090909) (0.0903342366757001,0.939393939393939) (0.178139114724481,0.96969696969697) (0.261427280939476,0.96969696969697) (0.263775971093044,0.96969696969697) (0.26576332429991,0.96969696969697) (0.347967479674797,0.96969696969697) (0.422764227642276,0.96969696969697) (0.498644986449864,0.96969696969697) (0.573441734417344,0.96969696969697) (0.647154471544715,1) (0.648238482384824,1) (0.720686540198735,1) (0.78970189701897,1) (0.86070460704607,1) (0.931526648599819,1) (1,1)
  };
 \addplot[mark=none, color=green, dashed] coordinates { (0.022583559168925,0.696969696969697) (0.0726287262872629,0.818181818181818) (0.0789521228545619,0.818181818181818) (0.0829268292682927,0.848484848484849) (0.0861788617886179,0.848484848484849) (0.0903342366757001,0.848484848484849) (0.178139114724481,0.909090909090909) (0.261427280939476,0.909090909090909) (0.263775971093044,0.909090909090909) (0.26576332429991,0.909090909090909) (0.347967479674797,0.909090909090909) (0.422764227642276,0.939393939393939) (0.498644986449864,0.939393939393939) (0.573441734417344,0.939393939393939) (0.647154471544715,0.939393939393939) (0.648238482384824,0.939393939393939) (0.720686540198735,0.96969696969697) (0.78970189701897,0.96969696969697) (0.86070460704607,0.96969696969697) (0.931526648599819,0.96969696969697) (1,0.96969696969697)
  };

 \addplot[mark=none, color=red        ] coordinates { (0.000696320641729103,0.0397350993377483) (0.00488817090493831,0.165562913907285) (0.01141269531794,0.198675496688742) (0.0221708492326547,0.258278145695364) (0.0356864328886166,0.437086092715232) (0.0470921650001393,0.629139072847682) (0.0537907695735732,0.682119205298013) (0.0591245856892182,0.768211920529801) (0.0634487368743559,0.774834437086093) (0.0660251232487536,0.794701986754967) (0.0680166002840988,0.80794701986755) (0.0700637829707824,0.80794701986755) (0.0716235412082556,0.80794701986755) (0.0731902626521461,0.80794701986755) (0.0744993454585968,0.80794701986755) (0.0758084282650475,0.80794701986755) (0.0773403336768515,0.80794701986755) (0.0790393560426705,0.80794701986755) (0.0804877029774671,0.814569536423841) (0.082207614962538,0.821192052980132) (0.0838509316770186,0.821192052980132) (0.0854594323594129,0.821192052980132) (0.0868799264685402,0.821192052980132) (0.088787845026878,0.821192052980132) (0.0904450881541932,0.827814569536424) (0.0919352143274935,0.827814569536424) (0.093501935771384,0.827814569536424) (0.0951731053115338,0.827814569536424) (0.0967189371361725,0.827814569536424) (0.0982578057543938,0.834437086092715) (0.099650447037852,0.834437086092715) (0.101036125114893,0.834437086092715) (0.102519288081776,0.834437086092715) (0.103842297301061,0.834437086092715) (0.10554131966688,0.834437086092715) (0.106982703395259,0.834437086092715) (0.108055037183522,0.834437086092715) (0.109259671893714,0.834437086092715) (0.110464306603905,0.834437086092715) (0.11189176391945,0.834437086092715) (0.113172993900231,0.834437086092715) (0.114544745564438,0.834437086092715) (0.115819012338802,0.834437086092715) (0.117225580035095,0.834437086092715) (0.118360582681113,0.834437086092715) (0.11937024761162,0.834437086092715) (0.120651477592402,0.834437086092715) (0.121932707573183,0.834437086092715) (0.123025930980698,0.834437086092715) (0.124509093947581,0.834437086092715) (0.125880845611787,0.834437086092715) (0.127036737877058,0.834437086092715) (0.128095145252486,0.834437086092715) (0.129007325293151,0.834437086092715) (0.130225886416177,0.834437086092715) (0.131569785254714,0.834437086092715) (0.132663008662229,0.834437086092715) (0.133630894354232,0.834437086092715) (0.134751970587416,0.847682119205298) (0.135845193994931,0.854304635761589) (0.136701668384258,0.854304635761589) (0.137446731470908,0.854304635761589) (0.138143052112637,0.854304635761589) (0.139090048185388,0.854304635761589) (0.139828148065621,0.860927152317881) (0.140677659248531,0.860927152317881) (0.141457538367267,0.867549668874172) (0.142251343898839,0.867549668874172) (0.143114781494583,0.867549668874172) (0.144075703980169,0.867549668874172) (0.145092332117093,0.867549668874172) (0.146081107428349,0.867549668874172) (0.146707796005905,0.867549668874172) (0.147522491156728,0.867549668874172) (0.148086510876528,0.867549668874172) (0.148949948472273,0.867549668874172) (0.149646269114002,0.867549668874172) (0.150495780296911,0.867549668874172) (0.151289585828482,0.867549668874172) (0.152146060217809,0.867549668874172) (0.153079129877726,0.874172185430464) (0.153845082583628,0.874172185430464) (0.154701556972955,0.874172185430464) (0.155620700220037,0.874172185430464) (0.15646324819653,0.874172185430464) (0.15720831128318,0.874172185430464) (0.158071748878924,0.874172185430464) (0.158928223268251,0.874172185430464) (0.159749881625491,0.874172185430464) (0.160724730523912,0.874172185430464) (0.161790101105757,0.874172185430464) (0.16286243489402,0.874172185430464) (0.163927805475866,0.874172185430464) (0.164895691167869,0.874172185430464) (0.165786981589282,0.874172185430464) (0.166685235217113,0.874172185430464) (0.167750605798958,0.874172185430464) (0.168384257582932,0.874172185430464) (0.169268584797928,0.874172185430464) (0.170055427123082,0.874172185430464) (0.17093279113166,0.874172185430464) (0.17193549285575,0.874172185430464) (0.172826783277163,0.894039735099338) (0.173745926524246,0.894039735099338) (0.174254240592708,0.894039735099338) (0.174873965963847,0.894039735099338) (0.175486728128569,0.894039735099338) (0.175862741275102,0.894039735099338) (0.176398908169234,0.894039735099338) (0.176705289251595,0.894039735099338) (0.17717878728797,0.894039735099338) (0.177478205163914,0.894039735099338) (0.17761746929226,0.894039735099338) (0.177889034342534,0.894039735099338) (0.178118820154305,0.894039735099338) (0.178334679553241,0.894039735099338) (0.178494833300838,0.894039735099338) (0.178703729493357,0.894039735099338) (0.178808177589616,0.894039735099338) (0.179051889814222,0.894039735099338) (0.179198117148985,0.894039735099338) (0.179295602038827,0.894039735099338) (0.17951842464418,0.894039735099338) (0.179755173662368,0.894039735099338) (0.179998885886973,0.894039735099338) (0.18011029718965,0.894039735099338) (0.197908252792246,0.900662251655629) (0.198994512993343,0.907284768211921) (0.199356599727042,0.927152317880795) (0.199565495919561,0.927152317880795) (0.199656017602986,0.927152317880795) (0.217621090159597,0.940397350993378) (0.218345263626995,0.947019867549669) (0.218937136172465,0.947019867549669) (0.219166921984235,0.947019867549669) (0.236902208729076,0.960264900662252) 
 (0.237605492577222,0.960264900662252) (0.237737793499151,0.960264900662252) (0.237849204801827,0.960264900662252) (0.237974542517338,0.960264900662252) (0.238085953820015,0.960264900662252) (0.238204328329109,0.960264900662252) (0.238343592457455,0.960264900662252) (0.238475893379383,0.960264900662252) (0.238559451856391,0.960264900662252) (0.238649973539816,0.960264900662252) (0.23874049522324,0.960264900662252) (0.2388449433195,0.960264900662252) (0.238942428209342,0.960264900662252) (0.256684678160599,0.966887417218543) (0.257074617719968,0.966887417218543) (0.257165139403392,0.966887417218543) (0.274768125226304,0.966887417218543) (0.275025763863744,0.966887417218543) (0.275137175166421,0.966887417218543) (0.275234660056263,0.966887417218543) (0.275332144946105,0.966887417218543) (0.292914241149765,0.966887417218543) (0.293074394897362,0.966887417218543) (0.293178842993622,0.966887417218543) (0.310642564688188,0.966887417218543) (0.311018577834721,0.980132450331126) (0.311123025930981,0.980132450331126) (0.328614600451216,0.980132450331126) (0.328851349469404,0.986754966887417) (0.328955797565663,0.986754966887417) (0.34631507116397,0.986754966887417) (0.346537893769323,0.993377483443709) (0.346774642787511,0.993377483443709) (0.346886054090187,0.993377483443709) (0.364419407848926,0.993377483443709) (0.364523855945186,0.993377483443709) (0.364635267247862,0.993377483443709) (0.364739715344122,0.993377483443709) (0.36487201626605,0.993377483443709) (0.364969501155892,0.993377483443709) (0.365060022839317,0.993377483443709) (0.382495891708214,0.993377483443709) (0.38260730301089,0.993377483443709) (0.38271175110715,0.993377483443709) (0.400161546388881,0.993377483443709) (0.400272957691558,0.993377483443709) (0.400377405787817,0.993377483443709) (0.400467927471242,0.993377483443709) (0.417868980308052,0.993377483443709) (0.417973428404312,0.993377483443709) (0.43550678216305,0.993377483443709) (0.435597303846475,0.993377483443709) (0.452991393476868,0.993377483443709) (0.453081915160293,0.993377483443709) (0.470364593488009,0.993377483443709) (0.470476004790686,0.993377483443709) (0.470566526474111,0.993377483443709) (0.470705790602457,0.993377483443709) (0.470810238698716,0.993377483443709) (0.470942539620645,0.993377483443709) (0.471026098097652,0.993377483443709) (0.471130546193911,0.993377483443709) (0.471269810322257,0.993377483443709) (0.471381221624934,0.993377483443709) (0.471499596134028,0.993377483443709) (0.47159708102387,0.993377483443709) (0.488907612177255,0.993377483443709) (0.506204216917806,0.993377483443709) (0.506308665014066,0.993377483443709) (0.523507784864775,0.993377483443709) (0.523598306548199,0.993377483443709) (0.523730607470128,0.993377483443709) (0.541110770687686,0.993377483443709) (0.541215218783946,0.993377483443709) (0.541326630086622,0.993377483443709) (0.54141018856363,0.993377483443709) (0.558692866891346,0.993377483443709) (0.558783388574771,0.993377483443709) (0.576156588585912,0.993377483443709) (0.576281926301423,0.993377483443709) (0.59353675180347,0.993377483443709) (0.610826393337604,0.993377483443709) (0.610909951814612,0.993377483443709) (0.628150850903824,0.993377483443709) (0.628255299000084,0.993377483443709) (0.645398713199454,0.993377483443709) (0.645503161295714,0.993377483443709) (0.645593682979138,0.993377483443709) (0.645719020694649,0.993377483443709) (0.645858284822995,0.993377483443709) (0.645990585744924,0.993377483443709) (0.663161852769964,0.993377483443709) (0.663287190485475,0.993377483443709) (0.680646464083781,0.993377483443709) (0.680750912180041,0.993377483443709) (0.680862323482717,0.993377483443709) (0.680987661198229,0.993377483443709) (0.681140851739409,0.993377483443709) (0.681259226248503,0.993377483443709) (0.681356711138345,0.993377483443709) (0.681454196028187,0.993377483443709) (0.698653315878896,0.993377483443709) (0.698757763975155,0.993377483443709) (0.716026515890037,0.993377483443709) (0.716130963986296,0.993377483443709) (0.71624933849539,0.993377483443709) (0.716367713004484,0.993377483443709) (0.716493050719996,0.993377483443709) (0.716625351641924,0.993377483443709) (0.716722836531766,0.993377483443709) (0.71684121104086,0.993377483443709) (0.716945659137119,0.993377483443709) (0.717084923265465,0.993377483443709) (0.73434671197393,0.993377483443709) (0.734458123276606,0.993377483443709) (0.734562571372866,0.993377483443709) (0.734667019469125,0.993377483443709) (0.734771467565384,0.993377483443709) (0.734889842074478,0.993377483443709) (0.734980363757903,0.993377483443709) (0.752214299640699,0.993377483443709) (0.769469125142746,0.993377483443709) (0.769629278890343,0.993377483443709) (0.787079074172075,0.993377483443709) (0.787559535414868,0.993377483443709) (0.78806784948333,0.993377483443709) (0.788569200345375,0.993377483443709) (0.788792022950728,0.993377483443709) (0.806401971980057,0.993377483443709) (0.806847617190764,0.993377483443709) (0.807348968052809,0.993377483443709) (0.807864245327689,0.993377483443709) (0.808114920758711,0.993377483443709) (0.825690053755954,0.993377483443709) (0.826198367824416,0.993377483443709) (0.84353675180347,0.993377483443709) (0.844010249839846,0.993377483443709) (0.844407152605632,1) (0.86152967718575,1) (0.878861097958388,1) (0.896324819652954,1) (0.913544829122915,1) (0.930785728212127,1) (0.948346934796535,1) (0.965706208394842,1) (0.982793916942874,1) (1,1)
  };
 \addplot[mark=none, color=red, dashed] coordinates { (0.000696320641729103,0.158940397350993) (0.00488817090493831,0.165562913907285) (0.01141269531794,0.172185430463576) (0.0221708492326547,0.178807947019868) (0.0356864328886166,0.205298013245033) (0.0470921650001393,0.225165562913907) (0.0537907695735732,0.218543046357616) (0.0591245856892182,0.211920529801325) (0.0634487368743559,0.225165562913907) (0.0660251232487536,0.231788079470199) (0.0680166002840988,0.23841059602649) (0.0700637829707824,0.23841059602649) (0.0716235412082556,0.23841059602649) (0.0731902626521461,0.23841059602649) (0.0744993454585968,0.23841059602649) (0.0758084282650475,0.23841059602649) (0.0773403336768515,0.23841059602649) (0.0790393560426705,0.23841059602649) (0.0804877029774671,0.231788079470199) (0.082207614962538,0.231788079470199) (0.0838509316770186,0.231788079470199) (0.0854594323594129,0.23841059602649) (0.0868799264685402,0.231788079470199) (0.088787845026878,0.231788079470199) (0.0904450881541932,0.231788079470199) (0.0919352143274935,0.231788079470199) (0.093501935771384,0.231788079470199) (0.0951731053115338,0.23841059602649) (0.0967189371361725,0.231788079470199) (0.0982578057543938,0.23841059602649) (0.099650447037852,0.231788079470199) (0.101036125114893,0.231788079470199) (0.102519288081776,0.231788079470199) (0.103842297301061,0.231788079470199) (0.10554131966688,0.231788079470199) (0.106982703395259,0.231788079470199) (0.108055037183522,0.231788079470199) (0.109259671893714,0.231788079470199) (0.110464306603905,0.231788079470199) (0.11189176391945,0.231788079470199) (0.113172993900231,0.231788079470199) (0.114544745564438,0.231788079470199) (0.115819012338802,0.231788079470199) (0.117225580035095,0.23841059602649) (0.118360582681113,0.23841059602649) (0.11937024761162,0.231788079470199) (0.120651477592402,0.231788079470199) (0.121932707573183,0.23841059602649) (0.123025930980698,0.23841059602649) (0.124509093947581,0.245033112582781) (0.125880845611787,0.245033112582781) (0.127036737877058,0.245033112582781) (0.128095145252486,0.245033112582781) (0.129007325293151,0.245033112582781) (0.130225886416177,0.245033112582781) (0.131569785254714,0.245033112582781) (0.132663008662229,0.245033112582781) (0.133630894354232,0.245033112582781) (0.134751970587416,0.245033112582781) (0.135845193994931,0.245033112582781) (0.136701668384258,0.245033112582781) (0.137446731470908,0.245033112582781) (0.138143052112637,0.245033112582781) (0.139090048185388,0.245033112582781) (0.139828148065621,0.245033112582781) (0.140677659248531,0.245033112582781) (0.141457538367267,0.245033112582781) (0.142251343898839,0.245033112582781) (0.143114781494583,0.245033112582781) (0.144075703980169,0.245033112582781) (0.145092332117093,0.245033112582781) (0.146081107428349,0.245033112582781) (0.146707796005905,0.245033112582781) (0.147522491156728,0.245033112582781) (0.148086510876528,0.245033112582781) (0.148949948472273,0.245033112582781) (0.149646269114002,0.245033112582781) (0.150495780296911,0.245033112582781) (0.151289585828482,0.251655629139073) (0.152146060217809,0.251655629139073) (0.153079129877726,0.251655629139073) (0.153845082583628,0.251655629139073) (0.154701556972955,0.251655629139073) (0.155620700220037,0.251655629139073) (0.15646324819653,0.251655629139073) (0.15720831128318,0.251655629139073) (0.158071748878924,0.251655629139073) (0.158928223268251,0.251655629139073) (0.159749881625491,0.251655629139073) (0.160724730523912,0.251655629139073) (0.161790101105757,0.251655629139073) (0.16286243489402,0.251655629139073) (0.163927805475866,0.245033112582781) (0.164895691167869,0.251655629139073) (0.165786981589282,0.251655629139073) (0.166685235217113,0.251655629139073) (0.167750605798958,0.251655629139073) (0.168384257582932,0.251655629139073) (0.169268584797928,0.251655629139073) (0.170055427123082,0.251655629139073) (0.17093279113166,0.251655629139073) (0.17193549285575,0.251655629139073) (0.172826783277163,0.251655629139073) (0.173745926524246,0.251655629139073) (0.174254240592708,0.251655629139073) (0.174873965963847,0.251655629139073) (0.175486728128569,0.251655629139073) (0.175862741275102,0.251655629139073) (0.176398908169234,0.251655629139073) (0.176705289251595,0.251655629139073) (0.17717878728797,0.251655629139073) (0.177478205163914,0.251655629139073) (0.17761746929226,0.251655629139073) (0.177889034342534,0.251655629139073) (0.178118820154305,0.251655629139073) (0.178334679553241,0.251655629139073) (0.178494833300838,0.251655629139073) (0.178703729493357,0.251655629139073) (0.178808177589616,0.251655629139073) (0.179051889814222,0.251655629139073) (0.179198117148985,0.251655629139073) (0.179295602038827,0.251655629139073) (0.17951842464418,0.251655629139073) (0.179755173662368,0.251655629139073) (0.179998885886973,0.251655629139073) (0.18011029718965,0.251655629139073) (0.197908252792246,0.258278145695364) (0.198994512993343,0.271523178807947) (0.199356599727042,0.271523178807947) (0.199565495919561,0.271523178807947) (0.199656017602986,0.271523178807947) (0.217621090159597,0.28476821192053) (0.218345263626995,0.28476821192053) (0.218937136172465,0.28476821192053) (0.219166921984235,0.28476821192053) (0.236902208729076,0.291390728476821) (0.237605492577222,0.291390728476821) (0.237737793499151,0.291390728476821) (0.237849204801827,0.291390728476821) (0.237974542517338,0.291390728476821) (0.238085953820015,0.291390728476821) 
 (0.238204328329109,0.291390728476821) (0.238343592457455,0.291390728476821) (0.238475893379383,0.291390728476821) (0.238559451856391,0.291390728476821) (0.238649973539816,0.291390728476821) (0.23874049522324,0.291390728476821) (0.2388449433195,0.291390728476821) (0.238942428209342,0.291390728476821) (0.256684678160599,0.28476821192053) (0.257074617719968,0.28476821192053) (0.257165139403392,0.28476821192053) (0.274768125226304,0.298013245033113) (0.275025763863744,0.298013245033113) (0.275137175166421,0.298013245033113) (0.275234660056263,0.298013245033113) (0.275332144946105,0.298013245033113) (0.292914241149765,0.311258278145695) (0.293074394897362,0.311258278145695) (0.293178842993622,0.311258278145695) (0.310642564688188,0.337748344370861) (0.311018577834721,0.337748344370861) (0.311123025930981,0.337748344370861) (0.328614600451216,0.324503311258278) (0.328851349469404,0.324503311258278) (0.328955797565663,0.324503311258278) (0.34631507116397,0.337748344370861) (0.346537893769323,0.337748344370861) (0.346774642787511,0.337748344370861) (0.346886054090187,0.337748344370861) (0.364419407848926,0.337748344370861) (0.364523855945186,0.337748344370861) (0.364635267247862,0.337748344370861) (0.364739715344122,0.337748344370861) (0.36487201626605,0.337748344370861) (0.364969501155892,0.337748344370861) (0.365060022839317,0.337748344370861) (0.382495891708214,0.33112582781457) (0.38260730301089,0.33112582781457) (0.38271175110715,0.33112582781457) (0.400161546388881,0.344370860927152) (0.400272957691558,0.344370860927152) (0.400377405787817,0.344370860927152) (0.400467927471242,0.344370860927152) (0.417868980308052,0.344370860927152) (0.417973428404312,0.344370860927152) (0.43550678216305,0.364238410596026) (0.435597303846475,0.364238410596026) (0.452991393476868,0.364238410596026) (0.453081915160293,0.364238410596026) (0.470364593488009,0.357615894039735) (0.470476004790686,0.357615894039735) (0.470566526474111,0.357615894039735) (0.470705790602457,0.357615894039735) (0.470810238698716,0.357615894039735) (0.470942539620645,0.357615894039735) (0.471026098097652,0.357615894039735) (0.471130546193911,0.357615894039735) (0.471269810322257,0.357615894039735) (0.471381221624934,0.357615894039735) (0.471499596134028,0.357615894039735) (0.47159708102387,0.357615894039735) (0.488907612177255,0.357615894039735) (0.506204216917806,0.377483443708609) (0.506308665014066,0.377483443708609) (0.523507784864775,0.390728476821192) (0.523598306548199,0.384105960264901) (0.523730607470128,0.384105960264901) (0.541110770687686,0.410596026490066) (0.541215218783946,0.410596026490066) (0.541326630086622,0.410596026490066) (0.54141018856363,0.410596026490066) (0.558692866891346,0.417218543046358) (0.558783388574771,0.417218543046358) (0.576156588585912,0.417218543046358) (0.576281926301423,0.417218543046358) (0.59353675180347,0.377483443708609) (0.610826393337604,0.397350993377483) (0.610909951814612,0.397350993377483) (0.628150850903824,0.423841059602649) (0.628255299000084,0.423841059602649) (0.645398713199454,0.437086092715232) (0.645503161295714,0.437086092715232) (0.645593682979138,0.437086092715232) (0.645719020694649,0.437086092715232) (0.645858284822995,0.437086092715232) (0.645990585744924,0.443708609271523) (0.663161852769964,0.443708609271523) (0.663287190485475,0.443708609271523) (0.680646464083781,0.456953642384106) (0.680750912180041,0.456953642384106) (0.680862323482717,0.456953642384106) (0.680987661198229,0.456953642384106) (0.681140851739409,0.456953642384106) (0.681259226248503,0.456953642384106) (0.681356711138345,0.456953642384106) (0.681454196028187,0.456953642384106) (0.698653315878896,0.456953642384106) (0.698757763975155,0.456953642384106) (0.716026515890037,0.470198675496689) (0.716130963986296,0.470198675496689) (0.71624933849539,0.470198675496689) (0.716367713004484,0.470198675496689) (0.716493050719996,0.470198675496689) (0.716625351641924,0.470198675496689) (0.716722836531766,0.470198675496689) (0.71684121104086,0.470198675496689) (0.716945659137119,0.470198675496689) (0.717084923265465,0.470198675496689) (0.73434671197393,0.463576158940397) (0.734458123276606,0.463576158940397) (0.734562571372866,0.463576158940397) (0.734667019469125,0.463576158940397) (0.734771467565384,0.463576158940397) (0.734889842074478,0.463576158940397) (0.734980363757903,0.463576158940397) (0.752214299640699,0.463576158940397) (0.769469125142746,0.456953642384106) (0.769629278890343,0.456953642384106) (0.787079074172075,0.456953642384106) (0.787559535414868,0.456953642384106) (0.78806784948333,0.456953642384106) (0.788569200345375,0.456953642384106) (0.788792022950728,0.456953642384106) (0.806401971980057,0.463576158940397) (0.806847617190764,0.456953642384106) (0.807348968052809,0.456953642384106) (0.807864245327689,0.456953642384106) (0.808114920758711,0.456953642384106) (0.825690053755954,0.456953642384106) (0.826198367824416,0.456953642384106) (0.84353675180347,0.463576158940397) (0.844010249839846,0.463576158940397) (0.844407152605632,0.463576158940397) (0.86152967718575,0.47682119205298) (0.878861097958388,0.490066225165563) (0.896324819652954,0.496688741721854) (0.913544829122915,0.490066225165563) (0.930785728212127,0.490066225165563) (0.948346934796535,0.483443708609272) (0.965706208394842,0.483443708609272) (0.982793916942874,0.496688741721854) (1,0.509933774834437)
  };

 \addplot[mark=none, color=black        ] coordinates { (0.0154738878143133,0.176470588235294) (0.0928433268858801,0.823529411764706) (0.111218568665377,1) (0.209864603481625,1) (0.308510638297872,1) (0.40715667311412,1) (0.505802707930368,1) (0.604448742746615,1) (0.70357833655706,1) (0.802224371373308,1) (0.901353965183752,1) (1,1)
  };
 \addplot[mark=none, color=black, dashed] coordinates { (0.0154738878143133,0.411764705882353) (0.0928433268858801,0.705882352941177) (0.111218568665377,0.705882352941177) (0.209864603481625,0.823529411764706) (0.308510638297872,0.941176470588235) (0.40715667311412,0.941176470588235) (0.505802707930368,0.941176470588235) (0.604448742746615,1) (0.70357833655706,1) (0.802224371373308,1) (0.901353965183752,1) (1,1)
   };

 \addplot[mark=none, color=brown        ] coordinates { (0.0173416669677373,0.261538461538462) (0.0447270493876224,0.830769230769231) (0.0536869106542866,0.846153846153846) (0.0595035947830485,0.876923076923077) (0.0647422233462192,0.876923076923077) (0.0679937859026699,0.876923076923077) (0.0708840637306261,0.907692307692308) (0.0748220672712164,0.907692307692308) (0.0779291159362694,0.907692307692308) (0.0815058347483652,0.907692307692308) (0.0838180570107302,0.907692307692308) (0.0855160952346544,0.907692307692308) (0.0863109216373424,0.907692307692308) (0.0873225188771271,0.907692307692308) (0.0887676577911052,0.907692307692308) (0.0905740814335778,0.907692307692308) (0.0937533870443296,0.907692307692308) (0.0976191336392211,0.907692307692308) (0.099606199645941,0.907692307692308) (0.100545539940027,0.907692307692308) (0.101918421908306,0.907692307692308) (0.104050001806424,0.907692307692308) (0.105495140720402,0.907692307692308) (0.107265435890025,0.907692307692308) (0.108638317858304,0.907692307692308) (0.109071859532498,0.907692307692308) (0.110155713717981,0.907692307692308) (0.110769897756422,0.907692307692308) (0.1122150366704,0.907692307692308) (0.112504064453196,0.907692307692308) (0.113696304057227,0.907692307692308) (0.114780158242711,0.907692307692308) (0.115358213808302,0.907692307692308) (0.116550453412334,0.907692307692308) (0.117814949962065,0.907692307692308) (0.118103977744861,0.907692307692308) (0.118645904837603,0.907692307692308) (0.119910401387333,0.907692307692308) (0.121608439611258,0.907692307692308) (0.122728422269591,0.907692307692308) (0.12345099172658,0.907692307692308) (0.124390332020665,0.907692307692308) (0.125979984826041,0.907692307692308) (0.173308284258824,0.907692307692308) (0.176198562086781,0.907692307692308) (0.177824343365006,0.907692307692308) (0.178980454496188,0.907692307692308) (0.179377867697532,0.907692307692308) (0.226742295603165,0.907692307692308) (0.229668701903971,0.907692307692308) (0.231619639437841,0.907692307692308) (0.232414465840529,0.907692307692308) (0.233173163770367,0.907692307692308) (0.280573720148849,0.907692307692308) (0.282452400737021,0.907692307692308) (0.282958199356913,0.907692307692308) (0.328263304310127,0.923076923076923) (0.330431012681094,0.923076923076923) (0.330900682828137,0.923076923076923) (0.377289641966834,0.938461538461538) (0.379276707973554,0.938461538461538) (0.424473427508219,0.938461538461538) (0.424906969182413,0.938461538461538) (0.469381119260089,0.938461538461538) (0.47125979984826,0.938461538461538) (0.472054626250948,0.938461538461538) (0.516203620072979,0.938461538461538) (0.560822284042054,0.938461538461538) (0.561328082661946,0.938461538461538) (0.60659705914231,0.938461538461538) (0.607572527909245,0.953846153846154) (0.608656382094729,0.953846153846154) (0.653058275226706,0.969230769230769) (0.696340185700351,0.969230769230769) (0.739513710755446,0.969230769230769) (0.783337548321832,0.984615384615385) (0.784023989305972,0.984615384615385) (0.82835362549225,0.984615384615385) (0.872032949167239,0.984615384615385) (0.915025831858087,0.984615384615385) (0.958199356913183,0.984615384615385) (1,0.984615384615385)
  };
 \addplot[mark=none, color=brown, dashed] coordinates { (0.0173416669677373,0.584615384615385) (0.0447270493876224,0.738461538461539) (0.0536869106542866,0.769230769230769) (0.0595035947830485,0.769230769230769) (0.0647422233462192,0.784615384615385) (0.0679937859026699,0.784615384615385) (0.0708840637306261,0.784615384615385) (0.0748220672712164,0.8) (0.0779291159362694,0.8) (0.0815058347483652,0.8) (0.0838180570107302,0.8) (0.0855160952346544,0.8) (0.0863109216373424,0.8) (0.0873225188771271,0.8) (0.0887676577911052,0.8) (0.0905740814335778,0.8) (0.0937533870443296,0.8) (0.0976191336392211,0.815384615384615) (0.099606199645941,0.815384615384615) (0.100545539940027,0.815384615384615) (0.101918421908306,0.815384615384615) (0.104050001806424,0.815384615384615) (0.105495140720402,0.815384615384615) (0.107265435890025,0.815384615384615) (0.108638317858304,0.830769230769231) (0.109071859532498,0.830769230769231) (0.110155713717981,0.830769230769231) (0.110769897756422,0.830769230769231) (0.1122150366704,0.830769230769231) (0.112504064453196,0.830769230769231) (0.113696304057227,0.830769230769231) (0.114780158242711,0.830769230769231) (0.115358213808302,0.830769230769231) (0.116550453412334,0.830769230769231) (0.117814949962065,0.830769230769231) (0.118103977744861,0.830769230769231) (0.118645904837603,0.815384615384615) (0.119910401387333,0.815384615384615) (0.121608439611258,0.861538461538462) (0.122728422269591,0.861538461538462) (0.12345099172658,0.861538461538462) (0.124390332020665,0.861538461538462) (0.125979984826041,0.861538461538462) (0.173308284258824,0.876923076923077) (0.176198562086781,0.876923076923077) (0.177824343365006,0.876923076923077) (0.178980454496188,0.876923076923077) (0.179377867697532,0.876923076923077) (0.226742295603165,0.892307692307692) (0.229668701903971,0.892307692307692) (0.231619639437841,0.892307692307692) (0.232414465840529,0.892307692307692) (0.233173163770367,0.892307692307692) (0.280573720148849,0.892307692307692) (0.282452400737021,0.892307692307692) (0.282958199356913,0.892307692307692) (0.328263304310127,0.907692307692308) (0.330431012681094,0.907692307692308) (0.330900682828137,0.907692307692308) (0.377289641966834,0.907692307692308) (0.379276707973554,0.907692307692308) (0.424473427508219,0.907692307692308) (0.424906969182413,0.907692307692308) (0.469381119260089,0.907692307692308) (0.47125979984826,0.907692307692308) (0.472054626250948,0.907692307692308) (0.516203620072979,0.907692307692308) (0.560822284042054,0.907692307692308) (0.561328082661946,0.907692307692308) (0.60659705914231,0.907692307692308) (0.607572527909245,0.907692307692308) (0.608656382094729,0.907692307692308) (0.653058275226706,0.907692307692308) (0.696340185700351,0.907692307692308) (0.739513710755446,0.907692307692308) (0.783337548321832,0.907692307692308) (0.784023989305972,0.907692307692308) (0.82835362549225,0.907692307692308) (0.872032949167239,0.907692307692308) (0.915025831858087,0.907692307692308) (0.958199356913183,0.923076923076923) (1,0.923076923076923)
  };

 \legend{date, \emph{date}, join, \emph{join}, sort, \emph{sort}, grin, \emph{grin}, ngircd, \emph{ngircd}}
 \end{axis}
\end{tikzpicture}
\end{minipage}
\caption{Progress of {\tool} on generating interactions (italic \emph{program name} indicates randomized version)}
\label{fig:evolve}
\end{figure*}
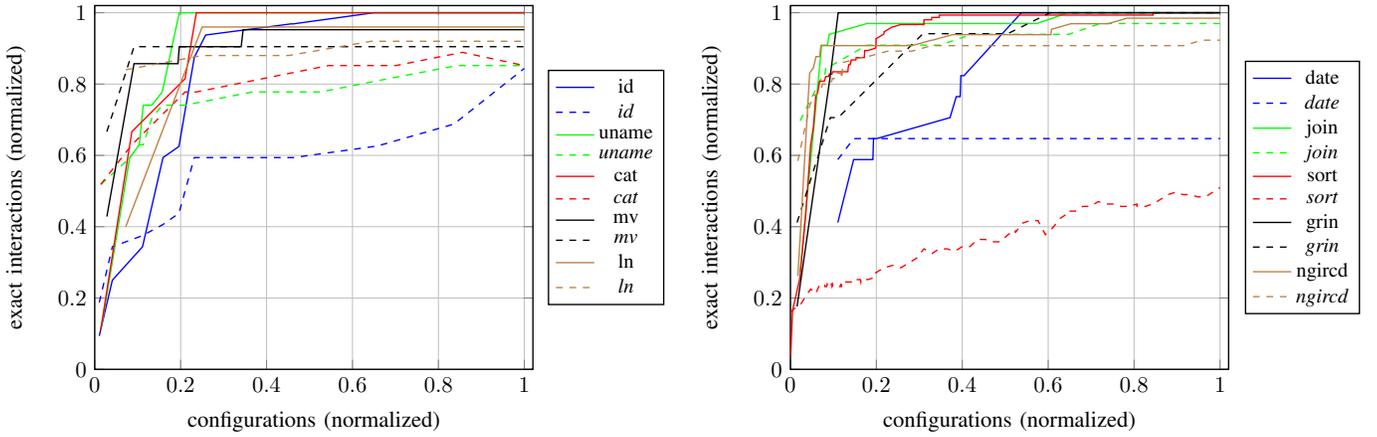

\paragraph*{Convergence}
Figure~\ref{fig:evolve} shows how {\tool} converges to its final results on the programs used in Table~\ref{tab:cmp}, which we can exhaustively run to obtain ground truth results.
The $x$-axis is the number of explored configurations (normalized such that 1 represents all configurations used by {\tool} for that particular program). The $y$-axis is the number of discovered interactions equivalent to ground truth (normalized such that 1 represents all interactions for that program).
These results show that {\tool} converges fairly quickly.
At around 40\% of configurations, {\tool} is able to accurately infer more than 90\% of the total ground truth interactions. It then spent the rest of the time refining few remaining difficult interactions.

\paragraph*{Comparing to Random Search}
We also compare interactions inferred from {\tool}'s configurations and randomly generated configurations.
For each program, we generate the same number of random configurations as the number of configurations {\tool} uses and then run {\myca} on these configurations to obtain interactions.

Figure~\ref{fig:evolve} shows that {\tool}'s configurations help the tool quickly outperform random configurations and stay dominated throughout the runs.
Comparing to random configurations, {\tool}'s configurations also learns more accurate interactions, especially for large programs or those with complex interactions, e.g., 
random configurations can only achieve about 56\% (84/151) of the ground truth interactions for \texttt{sort}.

\subsection{R3-Analysis}
\label{sec:analysis}

We analyze discovered interactions to learn interesting properties in configurable software. These experiments are  similar to those in previous interaction works~\cite{nguyen2016igen,reisner2010using,song:icse12, song:2014aa}.

\paragraph*{Interaction Forms}
Table~\ref{tab:results} shows that singular and conjunctive interactions are common, especially in small programs.
However, disjunctive interactions are relatively rare, e.g., 
only 1-2 disjunctions occur in the subject programs. %
Mixed interactions are also common, especially in large programs (e.g., in \texttt{sort}, \texttt{ls}, \texttt{unison}, and \texttt{bibtext2html}).
Existing works do not support many of these interactions and thus would not able to find them (see {\S}\ref{sec:cmp_igen}). %

\paragraph*{Interaction Length}
Table~\ref{tab:results} shows that the number of obtained interactions is far fewer than the number of possible interactions, which is consistent with prior works' results.
For example, for \texttt{id}, which has 10 boolean options, 1024 total configurations, and $2^{1024}$ possible interactions, {\tool} found only 32 interactions, which are many orders of magnitude less than $2^{1024}$.

Also, most interactions are relatively short, regardless of the number of configurations (e.g., all but \texttt{join}, \texttt{sort}, and \texttt{ack} have the median interaction lengths less than 10).
We also observe that we can achieve 74\% coverage using only interactions with length at most 3 and 93\% coverage with length at most 10. This observation is similar to previous works.

\paragraph*{Enabling Option}
Enabling options are those that must be set in a certain way to achieve significant coverage. For example, many locations in \texttt{coreutils} programs have interactions involving the conjunction $\overline{\texttt{help}} \land \overline{\texttt{version}}$.
Thus, both \texttt{help} and \texttt{version} are enabling options that must be turned off to reach those locations (because if either one is one, the program just prints a message and exits).
We also have the enabling options \texttt{Z} for \texttt{id} (because it is only applicable in SELinux-enabled kernel) and \texttt{ListenIPv4} for \texttt{ngircd} (this option need to be turned on to reach most of locations).
In general, enabling options are quite common, as suggested in previous works~\cite{reisner2010using,nguyen2016igen}.

\paragraph*{Minimal Covering Configurations}

A useful application of {\tool} is using the inferred interactions to compute a minimal set of configurations with high coverage.
To achieve this, we can use a greedy algorithm, e.g., the one described in iGen, which combines interactions having high coverage and no conflict settings, generates a configuration satisfying those interactions, and repeats this process until the generated configurations cover all interactions.

Column \texttt{min cspace} in Table~\ref{tab:results} shows that {\tool}'s interactions allow us to generate sets of high coverage configurations with sizes that are several orders of magnitude smaller than the sizes of configuration spaces.
For example, we only need 10/1024 configurations to cover 150 lines in \texttt{id} and 18/6291456 configurations to cover 1085 lines in \texttt{sort}.

\subsection{R4-Comparing to iGen}
\label{sec:cmp_igen}

Comparing to iGen, {\tool} generally explored more configurations but discovered more expressive interactions.
Table~\ref{tab:cmp}b compares the interactions inferred by {\tool} and iGen.
Column \texttt{pure} shows the number of single, purely conjunctive, and pure disjunctive interactions supported (and thus inferred) by both tools.
Columns \texttt{ok} and \texttt{fail} show the numbers of mixed interactions supported and not supported by iGen, respectively ({\tool} found all of these).
For example, both iGen and {\tool} discovered the purely conjunctive interaction  $\overline{\texttt{help}} \land \texttt{version} \land \overline{\texttt{Z}}$ for \texttt{id.c:182} and the mixed interaction $\overline{\texttt{help}} \land \overline{\texttt{version}} \land \overline{\texttt{Z}} \land \overline{\texttt{u}} \land (\overline{\texttt{g}} \lor \overline{\texttt{G}})$ for \texttt{id.c:198}.
However, only {\tool} inferred the more complex mixed interaction $\overline{\texttt{help}} \land \overline{\texttt{version}} \land \overline{\texttt{Z}} \land \overline{\texttt{g}} \land \overline{\texttt{G}} \land \overline{\texttt{n}} \land (\texttt{u} \lor (\overline{\texttt{r}} \land \overline{\texttt{z}}))$ for location \texttt{id.c:325}.

For small programs, we observe that many interactions are pure conjunctive or disjunctive, and hence, supported by both tools.
However, for larger and more complex programs (e.g., \texttt{sort}, \texttt{ngircd}), iGen could not generate most mixed interactions while {\tool} could.
For example, iGen failed to generate 122/132 of the mixed interactions in \texttt{sort} while {\tool} generated most of them.

\subsection{Threats to Validity}
Although the benchmark systems we have are popular and used in the real world, they only represent a small sample of configurable software systems.
Thus, our observations may not generalize in certain ways or to certain systems.
{\tool} runs the programs on test suites to obtains coverage information.
Our chosen tests have reasonable, but not complete, coverage.
Systems whose test suites are less (or more) complete could have different results. %
Our experiments used a substantial number of options, but do not include every possible configuration options.
We focused on subsets of configuration options that appeared to be important based on our experience.
Finally, {\tool} cannot infer interactions that cannot be represented by decision trees (e.g., configuration options involving non-finite numerical values).
Interactions involving such options might be important to the general understanding and analysis of configurable software.

\section{Related Work}
\label{sec:related}

\paragraph*{Interaction  Generation}
As mentioned, {\tool} is mostly related to iGen, which computes three forms of interactions: purely conjunctive, purely disjunctive, and specific mixtures of the two.
In contrast, we use decision trees to represent \emph{arbitrary} boolean interactions and develop our own classification algorithm {\myca} to manipulate decision trees.
{
\newcommand{\nottt}[1]{\overline{\texttt{#1}}}
\newcommand{\fhead}{\nottt{help} \land \nottt{version} \land \nottt{Z} \land \nottt{g} \land \nottt{G} \land \nottt{n}}
To illustrate the differences, consider the interaction for location \texttt{id.c:325}, $\fhead \land (\texttt{u} \lor (\nottt{r} \land \nottt{z}))$, which can be written as the disjunction of two purely conjunctive interactions: $(\fhead \land \texttt{u}) \vee (\fhead \land \nottt{r} \land \nottt{z})$.
iGen can infer each of these two purely conjunctions, but it \emph{cannot} discover their disjunction because iGen does not support this form, e.g., $(a \land b) \lor (a \land c)$. 
For this example, even when running on \emph{all} 1024 configurations, iGen only generates $\fhead \land \texttt{u}$, which misses the relation with $\texttt{r}$ and $\texttt{z}$.
In contrast, {\tool} generates this \emph{exact} disjunctive interaction (and many others) using 609 configurations in under a second (Table~\ref{tab:results} in {\S}\ref{sec:performance}).
}

Moreover, while both tools rely on the iterative guess-and-check approach, the learning and checking components and their integration in {\tool} are completely different from those in iGen, e.g., using heuristics to select likely fragile tree paths to generate counterexamples. Also, while {\myca} is a restricted case of C5.0, it is nonetheless a useful case that allows us to generate a tree that is exactly accurate over data instead of a tree that approximates the data.
We developed {\myca} because existing classification algorithms do not allow easy interaction inference (due to agressive pruning and simplification as explained in {\S}\ref{sec:myca}).

\paragraph*{Precondition and Invariant Discovery}
Researchers have used decision trees and general boolean formulae to represent program preconditions (interactions can be viewed as preconditions over configurable options).
The work in~\cite{sankaranarayanan2008dynamic} uses random SAT solving to generate data and decision trees to learn preconditions, but does not generate counterexample data to refine inferred preconditions, which we find crucial to improve resulting interactions.
Similarly, PIE~\cite{padhi2016data} uses PAC (probably approximately correct algorithm) to learn CNF formula over features to represent preconditions, but also does not generate counterexamples to validate or improve inferred results. Only when given the source code and postconditions to infer loop invariants PIE would be able to learn additional data using SMT solving.

{\tool} adopts the iterative refinement approach used in several invariant analyses (e.g., \cite{sharma2013data,garg2014ice,garg2016learning,nguyen2017syminfer}).
These works (in particular~\cite{garg2016learning,garg2014ice} that use decision trees)  rely on static analysis and constraint solving to check (and generate counterexamples) that the inferred invariants are correct with respect to the program with a given property/assertion (i.e., the purpose of these works is to prove correct programs correct). In contrast, {\tool} is pure dynamic analysis, in both learning and checking, and aims to discover interactions instead of proving certain goals.

\tool\ can be considered as a dynamic invariant tool that analyzes coverage trace information.
Daikon~\cite{ernst2001dynamically,ernst2007daikon} infers invariants from templates that fit program execution traces.
{\tool} focuses on inferring interactions represented by arbitrary formulae and combines with iterative refinement.
DySy is another invariant generator that uses symbolic execution for invariant inference~\cite{csallner2008dysy}.
The interaction work in~\cite{reisner2010using} also uses the symbolic executor Otter~\cite{ma2011directed} to fully explore the configuration space of a software system, but is limited to purely conjunctive formulae for efficiency.
Symbolic execution techniques often have similar limitations as static analysis, e.g., they require mocks or models to represent unknown libraries or frameworks and are language-specific (e.g., Otter only works on C programs).
Finally, \tool\ aims to discover new locations and learns interactions for all discovered locations.
In contrast, invariant generation tools typically consider a few specific locations (e.g., loop entrances and exit points).

\paragraph*{Binary decision diagrams (BDDs)}
The popular BDD data structure~\cite{akers1978binary} can be used to represent boolean formulae, and thus is an alternative to decision trees.
Two main advantages of BDDs are that a BDD can compactly represent a large decision tree and equivalent formulae are represented by the same BDD, which is desirable for equivalence checking.

However, our priority is not to compactly represent interactions or check their equivalences, but instead to be able to infer interactions from a small set of data.
While {\myca} avoids aggressive prunings to improve accuracy, it is inherently a classification algorithm that computes results by generalizing training data (like the original C5.0 algorithm, {\tool} performs generalization by using heuristics to decide when to stop splitting nodes to build the tree as described in {\S}\ref{sec:myca}).
To create a BDD representing a desired interaction, we would need many configurations, e.g., $2^n+1$ \emph{miss} or $2^n-1$ \emph{hit} configurations to create a BDD for $a \land (b_1 \lor b_2 \lor \dots \lor b_n)$.
In contrast, {\myca} identifies and generalizes patterns from training data and thus require much fewer configurations.
For instance, the configuration space size of the example in Figure~\ref{fig:f1} is 3888, and from just 3 configurations $c_1,c_2,c_3$, {\myca} learns the interaction $\bar s$ because it sees that whenever $s \equiv 1$, $L8$ is \emph{miss}, and whenever $s \equiv 0$, $L8$ is \emph{hit}.
BDD would need 1944 configurations to infer the same interaction.

\paragraph*{Combinatorial Interaction Testing and Variability-Aware Analyses}
Combinatorial interaction testing (CIT)~\cite{cohen1996combinatorial,cohen2003constructing} is often used to find variability bugs in configurable systems.
One popular CIT approach is using $t$-way covering arrays to generate a set of configurations containing all $t$-way combinations of option settings at least once.
CIT is effective, but is expensive and requires the developers to choose $t$ a priori.
Thus developers will often set $t$ to small, causing higher strength interactions to be ignored.
{\tool} initializes its set of configurations using 1-way covering arrays.

Variability-Aware is another popular type of analysis to find variability bugs~\cite{lauenroth2009model,apel2010type,liebig2010analysis,thum2012analysis,kastner2012type,liebig2013scalable,mordahl2019empirical,apel2013exploring,apel2016feature,meinicke2016essential}.
\cite{thum2012analysis} classify problems in software product line research and surveys static analysis to solve them.
{\tool}'s interactions belong to the feature-based classification, and we propose a new dynamic analysis to analyze them.
\cite{apel2013exploring} study feature interactions in a system and their effects, including bug triggering, power consumption, etc.
{\tool} complements these results by analyzing interactions that affect code coverage.

\section{Conclusion}
We presented {\tool}, a new dynamic analysis technique to learn program interactions, which are formulae that describe the configurations covering a location.
{\tool} works by iteratively running a subject program under a test suite and set
of configurations; building decision trees from the resulting
coverage information; and then generating new configurations that aim to
refine the trees in the next iteration.
Experimental results show that {\tool} is effective in accurately finding complex interactions and scales well to large programs.

\section{Data Availability}
{\tool} and all benchmark data are available at the public Github repository~\cite{toolwebsite}. A snapshot of the tool and benchmark used in this paper is available at~\cite{toolsnapshot}.

\section*{Acknowledgment}
We thank the anonymous reviewers for helpful comments.
This work was supported in part by awards CCF-1948536 from the National Science Foundation and W911NF-19-1-0054 from the Army Research Office.
KimHao Nguyen is also supported by the UCARE Award from the University of Nebraska-Lincoln.

\newpage
\balance
\bibliographystyle{IEEEtran}
\bibliography{paper}
\end{document}